\documentclass[useAMS,usenatbib]{mn2e}

\usepackage{graphicx}
\usepackage{fixltx2e} % forces figure order to remain fixed
\usepackage{url}

\usepackage{amssymb,amsmath,multirow}
\usepackage[rel]{overpic}

\renewcommand{\epsilon}{\varepsilon}

\newcommand{\koi}{\mbox{KOI-54}}
\newcommand{\welsh}{W11}

\renewcommand{\vec}[1]{\boldsymbol{#1}}
\newcommand{\uvec}[1]{{{\hat{#1}}}}
\newcommand{\mr}[1]{\mathrm{#1}}

\def\simless{\mathbin{\lower 3pt\hbox
      {$\rlap{\raise 5pt\hbox{\propto}}\mathchar"7218$}}}
      
\newcommand{\Sun}{\odot}
\newcommand{\MSun}{M_\Sun}
\newcommand{\RSun}{R_\Sun}

\newcommand{\brunt}{Brunt-V\"{a}is\"{a}l\"{a}}
      
\newcommand{\del}{\nabla}
\newcommand{\delad}{\del_{\mr{ad}}}
\newcommand{\Ylm}{Y_{lm}}
\newcommand{\Hlam}{H_{\lambda m}^q}
\newcommand{\hans}{X_{lm}^k}
\newcommand{\hansnorm}{\widetilde{X}_{lm}^k}
\newcommand{\hanse}{X_{lm}^k(e)}

\newcommand{\eq}[1]{eq.~\eqref{#1}}
\newcommand{\eqp}[1]{eq.~\ref{#1}}

\newcommand{\ta}[1]{Table~\ref{#1}}
\newcommand{\fig}[1]{Fig.~\ref{#1}}
\newcommand{\se}[1]{\S~\ref{#1}}
\newcommand{\ap}[1]{Appendix~\ref{#1}}

\newcommand{\obssub}{\mathrm{o}}

\newcommand{\Yo}{Y_{lm}(\theta_\obssub,\phi_\obssub)}

\newcommand{\Porb}{P_{\mr{orb}}}
\newcommand{\Prot}{P_{*}}
\newcommand{\Pps}{P_{\mr{ps}}}
\newcommand{\Ppse}{\Pps^\mr{nr}}
\newcommand{\Ppsp}{\Pps}

\newcommand{\Pbirth}{P_{\mr{birth}}}
\newcommand{\Pa}{P_1}
\newcommand{\Pb}{P_2}

\newcommand{\omegorb}{\Omega_{\mr{orb}}}
\newcommand{\omegrot}{\Omega_*}
\newcommand{\omegps}{\Omega_{\mr{ps}}}
\newcommand{\omegpsp}{\omegps}
\newcommand{\omegpse}{\Omega_{\mr{ps}}^\mr{nr}}
\newcommand{\omegperi}{\Omega_{\mr{peri}}}
\newcommand{\omegbirth}{\Omega_{\mr{birth}}}
\newcommand{\omegdyn}{\omega_{\mr{dyn}}}
\newcommand{\omegnl}{\omega_{nl}}

\newcommand{\gamnl}{\gamma_{nl}}

\newcommand{\frot}{f_{*}}

\newcommand{\Dp}{D_\mr{peri}}

\newcommand{\dopp}{\sigma_{km}}

\newcommand{\incl}{i}

\newcommand{\sphsum}{\sum_{l=0}^\infty \sum_{m=-l}^l}
\newcommand{\qsphsum}{\sum_{l=2}^\infty \sum_{m=-l}^l}

\renewcommand{\deg}{{}^\circ}

\newcommand{\ts}{t_\mr{sync}}
\newcommand{\tc}{t_\mr{circ}}

\newcommand{\vsi}{v_\mr{rot}\sin \incl_*}

\newcommand{\overlap}{Q_{nl}}
\newcommand{\lorentz}{\Delta_{nlmk}}

\newcommand{\be}{\begin{equation}}
\newcommand{\ee}{\end{equation}}

\newcommand{\lp}{\left(}
\newcommand{\rp}{\right)}

\newcommand\tupstrut{\rule{0pt}{11pt}}
\newcommand\tdownstrut{\rule[-6pt]{0pt}{0pt}}

\newcommand{\acknowledgments}{\begin{small}
    \section*{Acknowledgments}\end{small}}
\newcommand\altaffilmark[1]{$^{#1}$}
\newcommand\altaffiltext[1]{$^{#1}$}
\title[Tidal asteroseismology]{Tidal asteroseismology: Kepler's KOI-54}

\author[Burkart, Quataert, Arras, and Weinberg]{
\parbox[t]{\textwidth}{
Joshua Burkart,\altaffilmark{1,2}
Eliot Quataert,\altaffilmark{1,2}
Phil Arras,\altaffilmark{3} and
Nevin N. Weinberg\altaffilmark{4}}
\vspace*{6pt} \\
\altaffiltext{1}{Department of Physics, 366 LeConte Hall, University of California,
Berkeley, CA\ \ 94720, USA}\\
\altaffiltext{2}{Department of Astronomy \& Theoretical Astrophysics
  Center, 601 Campbell Hall, University of California Berkeley, CA\ \ 94720, USA}\\
\altaffiltext{3}{Department of Astronomy, University of Virginia,
P.O. Box 400325, Charlottesville, VA\ \ 22904-4325, USA}\\
\altaffiltext{4}{Department of Physics and MIT Kavli Institute, MIT, 77 Massachusetts Avenue,
Cambridge, MA~~02139, USA}
}

\date{Submitted to MNRAS August 18th, 2011}
\begin{document}
\maketitle
\label{firstpage}

\nonfrenchspacing
\begin{abstract}
We develop a general framework for interpreting and analyzing high-precision lightcurves from eccentric stellar binaries. Although our methods are general, we focus on the recently discovered Kepler system KOI-54, a face-on binary of two A stars with $e=0.83$ and an orbital period of 42 days. KOI-54 exhibits strong ellipsoidal variability during its periastron passage; its lightcurve also contains $\sim$ 20 pulsations at perfect harmonics of the orbital frequency, and another $\sim$ 10 nonharmonic pulsations. Analysis of such data is a new form of asteroseismology in which oscillation amplitudes and phases rather than frequencies contain information that can be mined to constrain stellar properties. We qualitatively explain the physics of mode excitation and the range of harmonics expected to be observed. To quantitatively model observed pulsation spectra, we develop and apply a linear, tidally forced, nonadiabatic stellar oscillation formalism including the Coriolis force. We produce temporal power spectra for KOI-54 that are semi-quantitatively consistent with the observations. Both stars in the KOI-54 system are expected to be rotating pseudosynchronously, with resonant nonaxisymmetric modes providing a key contribution to the total torque; such resonances present a possible explanation for the two largest-amplitude harmonic pulsations observed in KOI-54, although we find problems with this interpretation. We show in detail that the nonharmonic pulsations observed in KOI-54 can be explained by nonlinear three-mode coupling. The methods developed in this paper can be generalized in the future to determine the best-fit stellar parameters given pulsation data. We also derive an analytic model of KOI-54's ellipsoidal variability, including both tidal distortion and stellar irradiation, which can be used to model other similar systems.
\end{abstract}


\section{Introduction}
\label{s:introduction}
The recently discovered Kepler system \koi{} (\citealt{welsh11}; henceforth \welsh) is a highly eccentric stellar binary with a striking lightcurve: a 20-hour 0.6\% brightening occurs with a periodicity of 41.8 days, with lower-amplitude perfectly sinusoidal oscillations occurring in between. Such observations were only possible due to the unprecedented photometric precision afforded by Kepler. \welsh{} arrived at the following interpretation of these phenomena: during the periastron passage of the binary, each of its two similar A stars is maximally subjected to both its companion's tidal force and radiation field. The tidal force causes a prolate ellipsoidal distortion of each star known as the equilibrium tide, so that the resulting perturbations to both the stellar cross section and the emitted stellar flux produce a change in the observed flux. Along with the effects of irradiation, this then creates the large brightening at periastron. Secondly, the strong tidal force also resonantly excites stellar eigenmodes during periastron, which continue to oscillate throughout the binary's orbit due to their long damping times; this resonant response is known as the dynamical tide.

\welsh{} successfully exploited \koi's periastron flux variations, known traditionally as ellipsoidal variability, by optimizing a detailed model against this component of \koi's lightcurve \citep{orosz00}. In this way, \welsh{} were able to produce much tighter constraints on stellar and orbital parameters than could be inferred through traditional spectroscopic methods alone. \welsh{} also provided data on the dynamical tide oscillations. Thirty such pulsations were reported, of which roughly two-thirds have frequencies at exact harmonics of the orbital frequency. It is the analysis of these and similar future data that forms the basis of our work.

In close binary systems, tides provide a key mechanism to circularize orbits and synchronize stellar rotation with orbital motion. An extensive literature exists on the theory of stellar tides (e.g.,\ \citealt{zahn75,goodman98,witte99}). We have synthesized this theoretical formalism, together with other aspects of stellar oscillation theory, in order to model the dynamical tide of \koi{} as well as to provide a framework for interpreting other similar systems.

The methods we have begun to develop are a new form of asteroseismology, a long-standing subject with broad utility. In traditional asteroseismology, we observe stars in which internal stellar processes (e.g., turbulent convection or the kappa mechanism) drive stellar eigenmodes, allowing them to achieve large amplitudes \citep{dalsgaardnotes}. In this scenario, modes ring at their natural frequencies irrespective of the excitation mechanism. The observed frequencies (and linewidths) thus constitute the key information in traditional asteroseismology, and an extensive set of theoretical techniques exist to invert such data in order to infer stellar parameters and probe different aspects of stellar structure \citep{unno89}.

In tidal asteroseismology of systems like \koi, however, we observe modes excited by a periodic tidal potential from an eccentric orbit; tidal excitation occurs predominantly at $l=2$ (\se{s:nmodes}). Since orbital periods are well below a star's dynamical timescale, it is g-modes (buoyancy waves) rather than higher-frequency p-modes (sound waves) that primarily concern us. Furthermore, since modes in our case are \emph{forced} oscillators, they do not ring at their natural eigenfrequencies, but instead at pure harmonics of the orbital frequency. (We discuss nonharmonic pulsations in \se{s:anom}.) It is thus pulsation amplitudes and phases that provide the key data in tidal asteroseismology.

This set of harmonic amplitudes and phases in principle contains a large amount of information. One of the goals for future study is to determine exactly how the amplitudes can be optimally used to constrain stellar properties, e.g., the radial profile of the \brunt{} frequency. In this work, however, we focus on the more modest tasks of delineating the physical mechanisms at work in eccentric binaries and constructing a coherent theoretical model and corresponding numerical method capable of quantitatively modeling their dynamical tidal pulsations.

This paper is organized as follows. In \se{s:koibg} we give essential background on \koi. In \se{s:background} we give various theoretical results that we rely on in later sections, including background on tidal excitation of stellar eigenmodes (\se{s:nmodes}), techniques for computing disk-averaged observed flux perturbations (\se{s:obsflux}), and background on including the Coriolis force using the traditional approximation (\se{s:rot}). In \se{s:qualres} we use these results to qualitatively explain the pulsation spectra of eccentric stellar binaries, particularly what governs the range of harmonics excited. In \se{s:sync} we confront the rotational evolution of \koi's stars, showing that they are expected to have achieved a state of stochastic pseudosynchronization.

In \se{s:results} we present the results of our more detailed modeling. This includes an analytic model of ellipsoidal variability (\se{s:ellips}), the effects of nonadiabaticity (\se{s:nonad}), the effects of fast rotation  (\se{s:rotres}), and a preliminary optimization of our nonadiabatic method against \koi's pulsation data (\se{s:model}). We show in \se{s:anom} that the observed nonharmonic pulsations in \koi{} are well explained by nonlinear three-mode coupling, and perform estimates of instability thresholds, which may limit the amplitudes modes can attain.  We also address whether the highest-amplitude observed harmonics in \koi{} are signatures of resonant synchronization locks in \se{s:psmode}. We present our conclusions and prospects for future work in \se{s:conc}.

A few weeks prior to the completion of this manuscript, we became aware of a complementary study of \koi's pulsations \citep{fuller11}.

\section{Background on \koi{}}\label{s:koibg}
\ta{t:wparams} gives various parameters for \koi{} resulting from \welsh's observations and modeling efforts. \ta{t:data} gives a list of the pulsations \welsh{} reported, including both frequencies and amplitudes.

\begin{table}
\caption{List of \koi{} system parameters as determined by \welsh. Selected from Table 2 of \welsh. The top rows contain standard observables from stellar spectroscopy, whereas the bottom rows result from \welsh's modeling of photometric and radial velocity data. Symbols either have their conventional definitions, or are defined in \se{s:prelim}. Note that \welsh's convention is to use the less-massive star as the primary.}
\label{t:wparams}
\centering
\begin{tabular}{llrrc}\cline{2-5}
&\tupstrut \tdownstrut parameter & value& error & unit\\\cline{2-5}
\multirow{7}{*}{\rotatebox{90}{Observations}}&\tupstrut $T_{1}$     & 8500  & 200  & K \\
&$T_{2}$     & 8800  & 200  & K \\
&$L_2/L_1$ & 1.22 & 0.04 & \\ 
&$v_{\mr{rot},1} \sin{i_1}$ & 7.5  & 4.5 & $\mr{km}/\mr{s}$ \\
&$v_{\mr{rot},2} \sin{i_2}$ & 7.5  & 4.5 & $\mr{km}/\mr{s}$ \\
&$[\text{Fe}/\text{H}]_1$  & 0.4  & 0.2  &   \\
&\tdownstrut$[\text{Fe}/\text{H}]_2$  & 0.4  & 0.2  &   \\
\cline{2-5}
\multirow{10}{*}{\rotatebox{90}{Lightcurve/RV modeling}}& \tupstrut $M_{2}/M_{1}$     & 1.025    & 0.013   &      \\
&$\Porb$                 & 41.8051  & 0.0003  & days \\%
&$e$           & 0.8342   & 0.0005  &      \\
&$\omega$         & 36.22    & 0.90    & degrees \\
&$\incl$            & 5.52     & 0.10    & degrees \\
&$a$                 & 0.395    & 0.008   & AU    \\
&$M_{1}$          & 2.32   & 0.10  & $\MSun$ \\
&$M_{2}$          & 2.38   & 0.12  & $\MSun$ \\
&$R_{1}$        & 2.19   & 0.03  & $\RSun$ \\
&$R_{2}$        & 2.33   & 0.03  & $\RSun$\tdownstrut \\\cline{2-5}
\end{tabular}\rule{20pt}{0pt}
\end{table}

\begin{table}
\caption{Thirty largest KOI-54 pulsations. Originally Table 3 from \welsh. Asterisks ($*$) denote pulsations which are not obvious harmonics of the orbital frequency.}
\label{t:data}
\centering
\begin{tabular}{@{~~~}lr@{.}lc@{}l@{}}\hline
ID & \multicolumn{2}{c}{amp.\ ($\mu\mr{mag}$)} & $\omega/\omegorb$ \\\hline
F1  & \phantom{5551}297&7 &  90.00    \\
F2  & 229&4 &  91.00 \\
F3  &  97&2 &  44.00    \\
F4  &  82&9 &  40.00  \\
F5  &  82&9 &  22.42 & $\ *$\\
F6  &  49&3 &  68.58 & $\ *$  \\
F7  &  30&2 &  72.00   \\
F8  &  17&3 &  63.07 & $\ *$   \\
F9  &  15&9 &  57.58 & $\ *$  \\
F10 &  14&6 &  28.00  \\
F11 &  13&6 &  53.00   \\
F12 &  13&4 &  46.99   \\
F13 &  12&5 &  39.00    \\
F14 &  11&6 &  59.99   \\
F15 &  11&5 &  37.00   \\
F16 &  11&4 &  71.00    \\
F17 &  11&1 &  25.85 & $\ *$   \\
F18 &   9&8 &  75.99    \\
F19 &   9&3 &  35.84 & $\ *$  \\
F20 &   9&1 &  27.00  \\
F21 &   8&4 &  42.99   \\
F22 &   8&3 &  45.01   \\
F23 &   8&1 &  63.09 & $\ *$  \\
F24 &   6&9 &  35.99  \\
F25 &   6&8 &  60.42 & $\ *$   \\
F26 &   6&4 &  52.00   \\
F27 &   6&3 &  42.13  & $\ *$  \\
F28 &   5&9 &  33.00    \\
F29 &   5&8 &  29.00   \\
F30 &   5&7 &  48.00   \\\hline
\end{tabular}
\end{table}

\subsection{Initial rotation}\label{s:irot}
\koi's two components are inferred to be A stars. Isolated A stars are observed to rotate much more rapidly than e.g.\ the Sun, with typical surface velocities of $\sim$ 100 km/s and rotation periods of $\sim$ 1 day \citep{adelman04}. This results from their lack of a significant convective envelope, which means they experience less-significant magnetic braking, allowing them to retain more of their initial angular momentum as they evolve onto the main sequence. We thus operate under the assumption that both component stars of \koi{} were born with rotation periods of roughly
\[
\Pbirth \approx 1\ \mr{day}.
\]

\subsection{Rotational inclination}\label{s:obsrot}
\welsh{} constrained both stars' rotation, via line broadening, to be $\vsi = 7.5\pm 4.5$ km/s (the same for both stars). Using the mean values of $R_1$ and $R_2$ obtained from \welsh{}'s modeling, we can translate this into the following constraints on rotation periods (in days):
\[ \begin{split}
9.2 < \Pa/\sin \incl_1 < 37\qquad\text{and}\qquad
9.8 < \Pb/\sin \incl_2 < 39,
\end{split} \]
where $(\incl_1, \incl_2)$ and $(\Pa, \Pb)$ are the rotational angular momentum inclinations with respect to the observer and the stellar rotation periods, respectively.  If we assume that tidal interactions cause both stellar rotation periods to be approximately equal to the pseudosynchronous period of $\Ppsp \sim 1.8$ days derived in \se{s:sync}, we can constrain $\incl_1$ and $\incl_2$:
\[
2.8\deg < \incl_1 < 11\deg\qquad\mr{and}\qquad 2.6\deg < \incl_2 < 11\deg.
\]

\welsh{} obtained $\incl_\mr{orb} = 5.52 \deg \pm 0.10$ by fitting the lightcurve's ellipsoidal variation together with radial velocity measurements, so the constraints just derived are consistent with alignment of rotational and orbital angular momenta,
\begin{equation}\label{e:align}
\incl=\incl_\mr{orb}=\incl_1=\incl_2.
\end{equation}
Tides act to drive these three inclinations to be mutually parallel or antiparallel, so such an alignment once achieved is expected to persist indefinitely. In order to simplify the analytical formalism as well as reduce the computational expense of modeling the observed pulsations, we will adopt \eq{e:align} as an assumption for the rest of our analysis.

\section{Theoretical Background}
\label{s:background}
In this section we review various heterogeneous theoretical results that we rely on in later sections. In \se{s:prelim} we summarize the conventions and definitions used in our analysis. In \se{s:nmodes} we review the theory of tidally forced adiabatic stellar eigenmodes. Later (\se{s:qualres}), we use this formalism to explain qualitative features of the lightcurves of eccentric binaries like \koi.   We also use adiabatic normal modes to compute tidal torques (\se{s:sync} and \ap{a:sync}), as well as to perform a nonlinear saturation calculation (\se{s:anom}). However, our detailed quantitative modeling of the observations of \koi{} utilizes a nonadiabatic tidally forced stellar oscillation method that we introduce and employ in \se{s:results}. 

In \se{s:obsflux} we summarize how perturbed quantities at the stellar photosphere, specifically the radial displacement and Lagrangian flux perturbation, can be averaged over the stellar disk and translated into an observed flux variation. Lastly, in \se{s:rot} we review the traditional approximation, a way of simplifying the stellar oscillation equations in the presence of rapid rotation.

\subsection{Conventions and definitions}\label{s:prelim}
We label the two stars as per \ta{t:wparams}, consistent with \welsh{}; note that the primary/star 1 is taken to be the smaller and less massive star. In the following, we focus our analysis on star 1, since the results for star 2 are similar. We assume that both stars' rotational angular momentum vectors are perpendicular to the orbital plane (\se{s:obsrot}), and work in spherical coordinates $(r,\theta,\phi)$ centered on star 1 where $\theta=0$ aligns with the system's orbital angular momentum and $\phi=0$ points from star 1 to star 2 at periastron.

We write the stellar separation as  $D(t)$ and the true anomaly as $f(t)$, so that the position of star 2 is $\vec{D}=(D,\pi/2,f)$. We write the semi-major axis as $a$ and the eccentricity as $e$.   The angular position of the observer in these coordinates is $\hat{n}_o=(\theta_o,\phi_o)$, where these angles are related to the traditional inclination $i$ and argument of periastron $\omega$ by \citep{arras11}
\begin{equation}\label{e:obsang}
\theta_o=i\qquad\mr{and}\qquad \phi_o = \frac{\pi}{2} - \omega \mod 2\pi.
\end{equation}

The orbital period [angular frequency] is $\Porb$ [$\omegorb$], while a rotation period [angular frequency] is $\Prot$ [$\omegrot$]. The effective orbital frequency at periastron is
\begin{equation}
 \omegperi = \left.\frac{df}{dt}\right|_{f=0} = \frac{\omegorb}{1-e}\sqrt{\frac{1+e}{1-e}},
\end{equation}
which is $\omegperi=20.\times\omegorb$ for \koi. The stellar dynamical frequency is
\begin{equation}\label{e:dynf}
 \omegdyn = \sqrt{\frac{G M}{R^3}},
\end{equation} 
which is $\omegdyn \approx 1.1$ $\text{rad}/\text{hr}$ for \koi's stars.

\subsection{Tidal excitation of stellar eigenmodes}\label{s:nmodes}
Although we ultimately use an inhomogeneous, nonadiabatic code including the Coriolis force to model the pulsations in eccentric binaries (\se{s:nonad}), the well known normal mode formalism provides an excellent qualitative explanation for many of the features in the lightcurve power spectra of systems such as \koi. Here we will review the salient results of this standard theory;  we demonstrate their application to \koi{} and related systems in \S\se{s:qualres} -- \ref{s:sync}.  The remainder of the paper after \se{s:sync} primarily uses our nonadiabatic method described in \se{s:nonad}.

Working exclusively to linear order and operating in the coordinates specified in the previous section, we can represent the response of star 1 (and similarly for star 2)---all oscillation variables such as the radial displacement $\xi_r$, the Lagrangian pressure perturbation $\Delta p$, etc.---to a perturbing tidal potential by a spatial expansion in normal modes and a temporal expansion in orbital harmonics (e.g., \citealt{kumar95}):
\begin{equation}\label{e:summodes}
\delta X = \sum_{nlmk} A_{nlmk} \delta X_{nl}(r) e^{-ik\omegorb t}\Ylm(\theta, \phi).
\end{equation}
Here, $2\le l < \infty$ and $-l\le m\le l$ are the spherical harmonic quantum numbers, and index the angular expansion; $|n| < \infty$ is an eigenfunction's number of radial nodes, and indexes the radial expansion;\footnote{Conventionally, $n>0$ corresponds to p-modes while $n<0$ corresponds to g-modes; however, since we are mostly concerned with g-modes in this paper, we will report g-mode $n$ values as $>0$.} and $|k|< \infty$ is the orbital Fourier harmonic number, which indexes the temporal expansion.

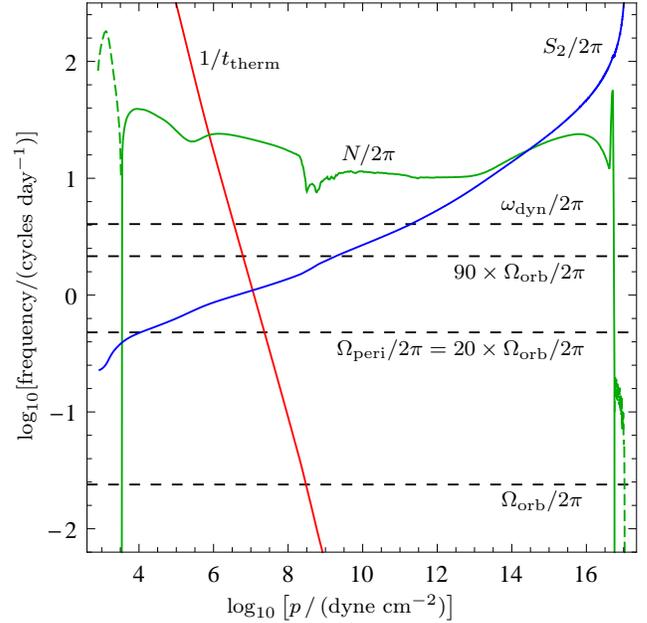
\begin{figure}
  \centering
  \begin{tabular}{@{}r@{\ }c}
    \rotatebox{90}{\rule{55pt}{0pt}$\log_{10}[\text{frequency} / (\text{cycles}\ \text{day}^{-1})]$}&
    \begin{overpic}{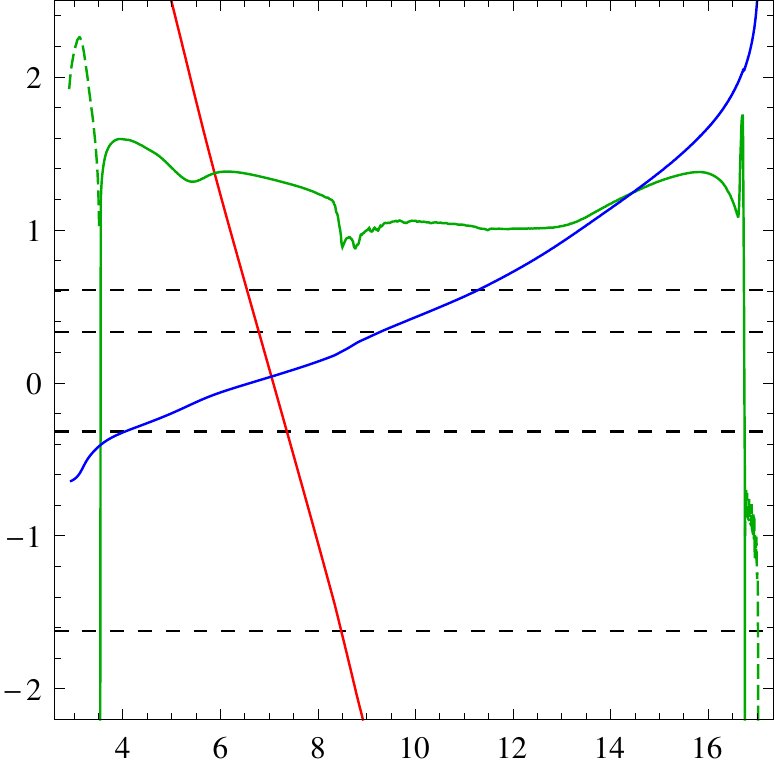}
    \put(50,72){$N/2\pi$}
    \put(84,90){$S_2/2\pi$}
    \put(26,88){$1/t_\mr{therm}$}
    \put(77,13.5){$\omegorb/2\pi$}
    \put(49.2,39){$\omegperi/2\pi=20\times\omegorb/2\pi$}
    \put(69,52){$90\times\omegorb/2\pi$}
    \put(77,63.5){$\omegdyn/2\pi$}
    \end{overpic}\\
    & $\log_{10}\big[p\,/\,(\text{dyne}\ \text{cm}^{-2})\big]$
  \end{tabular}
  \caption{Propagation diagram for a MESA stellar model consistent with \welsh's mean parameters for star 1 of \koi, showing the radial profiles of the \brunt{} frequency $N$, the Lamb frequency $S_l=\sqrt{l(l+1)}c_s/r$ for $l=2$ (where $c_s$ is the sound speed), and the inverse thermal time $1/t_\mr{therm}$ (defined in \eqp{e:ttherm}). ($N$ is dashed where $N^2<0$.) Propagation of g-modes occurs where the squared mode frequency $\omegnl^2$ is less than both $N^2$ and $S_l^2$ \citep{dalsgaardnotes}. Nonadiabatic effects become important when $\omegnl\cdot t_\mr{therm}<2\pi$. Several important frequencies for \koi{} are also plotted, which are defined in \se{s:prelim}. The 90th harmonic is the largest pulsation observed (\ta{t:data}).}
  \label{f:prop}
\end{figure}

Each $(n,l,m)$ pair formally corresponds to a distinct mode, although the eigenspectrum is degenerate in $m$ since for now we are ignoring the influence of rotation on the eigenmodes. Each mode has associated with it a set of eigenfunctions for the various perturbation variables, e.g., $\xi_r$, $\delta p$, etc., as well as a frequency $\omegnl$ and a damping rate $\gamma_{nl}$. For stars and modes of interest, $\gamma_{nl}$ is set by radiative diffusion; see the discussion after \eq{e:gamscale}. \fig{f:prop} gives a propagation diagram for a stellar model consistent with \welsh's mean parameters for star 1 (\ta{t:wparams}). The frequencies of g-modes behave asymptotically for $n\gg0$ and hence $\omegnl\ll\omegdyn$ as \citep{dalsgaardnotes}
\begin{equation}\label{e:gfreq}
 \omegnl\sim\omega_0 \frac{l}{n},
\end{equation}
where $\omega_0\approx 4$ $\text{rad}/\text{hr}$ for \koi's stars.

The amplitudes $A_{nlmk}$ appearing in \eq{e:summodes} each represent the pairing of a stellar eigenmode with an orbital harmonic. Their values are set by the tidal potential, and can be expressed analytically:
\begin{equation}\label{e:nmodes}
A_{nlmk} = \frac{2\,\epsilon_l \,\overlap\,\hansnorm \,W_{lm}\,\lorentz}{E_{nl}}.
\end{equation}
The coefficients appearing in \eq{e:nmodes} are as follows.
\begin{enumerate}
\item The tidal parameter $\epsilon_l$ is given by
\begin{equation}\label{e:tidefac}
\epsilon_l = \left(\frac{M_2}{M_1}\right)\left(\frac{R_1}{\Dp}\right)^{l+1},
\end{equation}
where $\Dp=a(1-e)$ is the binary separation at periastron. This factor represents the overall strength of the tide; due to its dependence on $R_1/\Dp$, which is a small number in cases of interest, it is often acceptable to consider only $l=2$.
\item The linear overlap integral $\overlap$ \citep{press77}, given by
\begin{equation}\begin{split}\label{e:overlap}
\overlap&=\frac{1}{M_1 R_1^l}\int_0^{R_1} l\left(\xi_{r,nl}+(l+1)\xi_{h,nl}\right)\rho r^{l+1} dr \\
&= \frac{1}{M_1 R_1^l}\int_0^{R_1} \delta\rho_{nl}\, r^{l+2} dr\\
& = -\frac{R_1}{GM_1}\cdot\frac{2l+1}{4\pi}\cdot\delta\phi(R_1),
\end{split}\end{equation}
represents the spatial coupling of the tidal potential to a given eigenmode; it is largest for modes with low $|n|$ and hence for eigenfrequencies close to the dynamical frequency $\omegdyn=\sqrt{G M_1/R_1^3}$, but falls off as a power law for $|n|\gg0$.
\item We define our mode normalization/energy $E_{nl}$ as
\begin{equation}
E_{nl} = 2\left(\frac{\omegnl^2R_1}{G M_1^2}\right)\int_0^{R_1} \left(\xi_{r,nl}^2+l(l+1)\xi_{h,nl}^2\right)\rho r^2 dr,
\end{equation}
where $\xi_h$ is the horizontal displacement \citep{dalsgaardnotes}.
\item The unit-normalized Hansen coefficients $\hansnorm$ are the Fourier series expansion of the orbital motion \citep{murray99}, and are defined implicitly by
\begin{equation}\label{e:hans}
\left(\frac{\Dp}{D(t)}\right)^{l+1} e^{-imf(t)} = \sum_{k=-\infty}^\infty \hansnorm(e) e^{-i k \omegorb t}.
\end{equation}
They are related to the traditional Hansen coefficients $\hans$ by $\hans=\hansnorm/(1-e)^{l+1}$ and satisfy the sum rule 
\begin{equation}
 \sum_{k=-\infty}^\infty \hansnorm(e) = 1,
\end{equation}
which can be verified using \eq{e:hans}. (An explicit expression for $\hans$ is given in \eqp{e:hansexp}.) The Hansen coefficients represent the temporal coupling of the tidal potential to a given orbital harmonic. They peak near $k_\mr{peak}\sim m\omegperi/\omegorb$ but fall off exponentially for larger $|k|$.

\item The Lorentzian factor $\lorentz$ is
\begin{equation}\label{e:lorentz}
\lorentz = \frac{\omegnl^2}{\left(\omegnl^2-\dopp^2\right)-2i\gamma_{nl}\dopp}
\end{equation}
where $ \dopp = k\omegorb - m\omegrot$, and represents the temporal coupling of a given harmonic to a given mode. When its corresponding mode/harmonic pair approach resonance, i.e.~$\omegnl \approx \dopp$, $\lorentz$ can become very large; its maximum, for a perfect resonance, is half the simple harmonic oscillator quality factor, $q_{nl}$: $\Delta_\mr{pr} = i\omegnl/2\gamma_{nl} = iq_{nl}/2$.

\item $W_{lm}$ is defined in \eq{e:W} and represents the angular coupling of the tidal potential to a given mode; it is nonzero only for $\mr{mod}(l+m,2)=0$. In particular, $W_{2,\pm1}= 0$, meaning that $l=2$, $|m|=1$ modes are not excited by the tidal potential.

\item To calculate the quasiadiabatic damping rate $\gamma_{nl}$ within the adiabatic normal mode formalism,\footnote{We only use this approximate method of calculating damping rates when employing the adiabatic normal mode formalism; our nonadiabatic method introduced in \se{s:nonad} fully includes radiative diffusion.} we average the product of the thermal diffusivity $\chi$ with a mode's squared wavenumber $k^2$, weighted by the mode energy:
\begin{equation}\label{e:gamform}
  \gamma_{nl} = \frac{\int_0^{r_c} k^2\chi (\xi_r^2+l(l+1)\xi_h^2)\rho r^2 dr}{\int_0^{r_c} (\xi_r^2+l(l+1)\xi_h^2)\rho r^2 dr}, 
\end{equation}
where the thermal diffusivity $\chi$ is
\begin{equation}
 \chi = \frac{16\sigma T^3}{3\kappa \rho^2 c_p}.
\end{equation} 
The cutoff radius $r_c$ is determined by the minimum of the mode's outer turning point and the point where $\omegnl \cdot  t_\mr{therm} =2\pi$ (\citealt{dalsgaardnotes}), where the thermal time is
\begin{equation}\label{e:ttherm}
t_\mr{therm}=\frac{p c_p T}{gF}.
\end{equation}
When this cutoff is restricted by the mode period intersecting the thermal time, so that strong nonadiabatic effects are present inside the mode's propagation cavity, the mode becomes a traveling wave at the surface, and the standing wave/adiabatic normal mode approximation becomes less realistic. This begins to occur at a frequency (in the rest frame of the star) of $\sim50\times\omegorb$ for \koi, as can be seen in \fig{f:prop}. Fortunately, our calculations involving the normal mode formalism (\S\se{s:sync} \& \ref{s:anom}) center primarily on low-order modes that are firmly within the standing wave limit.

The g-mode damping rate scales roughly as
\begin{equation}\label{e:gamscale}
 \gamma_{nl}\sim \gamma_0\,n^s \sim \gamma_0 \left[l\left(\frac{\omega_0}{\omegnl}\right)\right]^s,
\end{equation} 
where we have used the asymptotic g-mode frequency scaling from \eq{e:gfreq}. In the standing wave regime, i.e.\ for $\omegnl\lesssim50\times\omegorb$, we find that $\gamma_0\sim 1\ \text{Myr}^{-1}$ and $s\sim4$. This large value for $s$ results from the fact that most of the damping occurs at the surface, and the cutoff radius is limited by the outer turning point where the mode frequency intersects the Lamb frequency. As the mode frequency declines, the cutoff radius moves outward toward smaller Lamb frequency and stronger damping, as can be seen in \fig{f:prop}. Without this behavior of the turning point, we would expect $s\sim 2$ since $k^2\propto n^2$ in \eq{e:gamform}.
\end{enumerate}

\subsection{Observed flux perturbation}\label{s:obsflux}
\label{s:bolo}
\newcommand{\bandpass}{\beta}
\newcommand{\bandpassT}{\bandpass(T)}
Throughout this work, perturbations to the emitted flux $\Delta F$ are understood to be bolometric, i.e.\ integrated over the entire electromagnetic spectrum. We correct for Kepler's bandpass to first order as follows. We define the bandpass correction coefficient $\bandpassT$ as the ratio of the bandpass-corrected flux perturbation $(\Delta F/F)_\mr{bpc}$ to the bolometric perturbation $(\Delta F/F)$, so that
\begin{equation}
 \left(\frac{\Delta F}{F}\right)_\mr{bpc} = \bandpassT \left(\frac{\Delta F}{F}\right).
\end{equation}

We assume Kepler is perfectly sensitive to the wavelength band $(\lambda_1,\lambda_2) = (400,865)$ nm \citep{koch10}, and is completely insensitive to all other wavelengths. Then $\bandpassT$ is given to first order by
\begin{equation}\label{e:bandpass}
 \bandpassT \approx \frac{\int_{\lambda_1}^{\lambda_2}(\partial B_\lambda/\partial \ln T) d\lambda}{4\int_{\lambda_1}^{\lambda_2}B_\lambda d\lambda},
\end{equation}
where $B_\lambda (T)$ is the Planck function. Using \welsh's mean parameters for \koi{} (\ta{t:wparams}), we have $\bandpass(T_1) = 0.81$ and $\bandpass(T_2) = 0.79$. Note that employing $\bandpass$ alone amounts to ignoring bandpass corrections due to limb darkening. We have also ignored the fact that in realistic atmospheres, the perturbed specific intensity depends on perturbations to gravity in addition to temperature; this is a small effect, however, as shown e.g.\ in \citet{robinson82}.

For completeness, we transcribe several results from \citet{pfahl08}, which allow a radial displacement field $\xi_r$ and a Lagrangian perturbation to the emitted flux $\Delta F$, both evaluated at the stellar surface, to be translated into a corresponding disk-averaged observed flux perturbation $\delta J$, as seen e.g.\ by a telescope \citep{dziembowski77}. While an emitted flux perturbation alters the observed flux directly, a radial displacement field contributes by perturbing a star's cross section.\footnote{A horizontal displacement field $\xi_h$ produces no net effect to first order---its influence cancels against perturbations to limb darkening, all of which is included in \eq{e:fluxvar}.}

Given $\xi_r$ and $\Delta F$ expanded in spherical harmonics as
\begin{align}\label{e:xiexpflux}
\xi_r &= \sphsum \xi_{r,lm}(t)\Ylm(\theta, \phi)\\\label{e:dFexpflux}
\Delta F &= \sphsum \Delta F_{lm}(t)\Ylm(\theta, \phi),
\end{align}
we can translate these into a fractional observed flux variation $\delta J/J$ to first order by
\begin{multline}\label{e:fluxvar}
\frac{\delta J}{J} = \sum_{l=0}^\infty \sum_{m=-l}^l\bigg[(2b_l-c_{l})\frac{\xi_{r,lm}(t)}{R}\\
+\bandpassT \,b_l\frac{\Delta F_{lm}(t)}{F(R)}\bigg]\Yo,
\end{multline}
where the disk-integral factors are
\begin{align}\label{e:diskb}
 b_l &= \int_0^1 \mu P_l(\mu)h(\mu)\,d\mu\\\label{e:diskc}
 c_{l} & = \int_0^1 \left[2\mu^2\frac{dP_l}{d\mu}-(\mu-\mu^3)\frac{d^2P_l}{d\mu^2}\right]h(\mu)\,d\mu,
\end{align}
$P_l(\mu)$ is a Legendre polynomial, and $h(\mu)$ is the limb darkening function, normalized as $\int_0^1\mu h(\mu)d\mu = 1$. For simplicity, we use Eddington limb darkening for all of our analysis, with $h(\mu) = 1+3\mu/2$; $b_l$ and $c_l$ in this case are given in \ta{t:diskint} for $0\leq l \leq 5$.

\begin{table}
\centering
\caption{First six disk-integral factors $b_l$ and $c_l$ from equations \eqref{e:diskb} and \eqref{e:diskc} for linear Eddington limb darkening, $h(\mu) = 1 +(3/2)\mu$.}\label{t:diskint}
\begin{tabular}{ccc}\hline
 $l$ & $b_l$ & $c_l$\\\hline
 0 & 1 & 0 \\
 1 & 17/24 & 17/12 \\
 2 & 13/40 & 39/20 \\
\hline
\end{tabular}
\begin{tabular}{ccc}\hline
 $l$ & $b_l$ & $c_l$\\\hline
 3 & 1/16 & 3/4 \\
 4 & -1/48 & -5/12 \\
 5 & -1/128 & -15/64\\
\hline
\end{tabular}
\end{table}

Since $Y_{2,\pm2}\propto \sin^2\theta$ and $Y_{2,0}\propto (3\cos^2\theta  - 1)$, and since \koi{} has $\theta_o=i\approx 5.5\deg$ (\ta{t:wparams}), \eq{e:fluxvar} shows that $m=\pm 2$ eigenmodes are a factor of $\sim$  200  less observable than $m=0$ modes. It is thus likely that nearly all of the observed pulsations in \koi{} have $m = 0$; the exceptions may be F1 and F2, as we discuss in \se{s:psmode}.

\subsection{Rotation in the traditional approximation}\label{s:rot}
Stellar rotation manifests itself in a star's corotating frame as the fictitious centrifugal and Coriolis forces \citep{unno89}. The centrifugal force directly affects the equilibrium structure of a star, which can then consequently affect stellar oscillations. Its importance, however, is characterized by $(\omegrot/\omegdyn)^2$, which is $\sim$  $10^{-2}$ for rotation periods and stellar parameters of interest here (\se{s:koibg}). As such we neglect rotational modification of the equilibrium stellar structure \citep{ipser90}.

The Coriolis force, on the other hand, affects stellar oscillations directly.
Given a frequency of oscillation $\sigma$, the influence of the Coriolis force is characterized by the dimensionless rotation parameter $q$ given by
\begin{equation}\label{e:qrot}
q=2\omegrot/\sigma,
\end{equation}
where large values of $|q|$ imply that rotation is an important effect that must be accounted for. Note that for simplicity we assume rigid-body rotation throughout. For the pulsations observed in \koi's lightcurve (\ta{t:data}), assuming both stars rotate at near the pseudosynchronous rotation period $\Ppsp\sim 1.8$ days discussed in \se{s:pseudo} and that $m=0$ (justified in the previous section), $q$ ranges from $0.5$ for $k=\sigma/\omegorb =90$ to $1.5$ for $k = 30$. Thus lower harmonics fall in the nonperturbative rotation regime, where rotation is a critical effect that must be fully included.

The ``traditional approximation'' \citep{chapman70} greatly simplifies the required analysis when the Coriolis force is included in the momentum equation. In the case of g-modes, it is applicable for
\be
1\gg\frac{2}{q}\cdot\frac{R}{H_p}\cdot\lp\frac{\omegrot}{|N|}\rp^2, \label{e:trad}
\ee
where $H_p=\rho g / p$ is the pressure scale height; outside of the convective cores of models we are concerned with in this work (where g-modes are evanescent anyway), \eq{e:trad} is well satisfied whenever rotation is significant. Here we will give a brief overview of the traditional approximation; we refer to \citet{bildsten96} for a more thorough discussion.

The traditional approximation changes the angular Laplacian, which occurs when deriving the nonrotating stellar oscillation equations, into the Laplace tidal operator $L^q_m$. (Without the traditional approximation, the oscillation equations for a rotating star are generally not separable.) It is thus necessary to perform the polar expansion of oscillation variables in eigenfunctions of $L^q_m$, known as the Hough functions $\Hlam(\mu)$ (where $\mu=\cos\theta$), rather than associated Legendre functions; the azimuthal expansion is still in $e^{im\phi}$. The eigenvalues of $L^q_m$ are denoted $\lambda$, and depend on $m$, the azimuthal wavenumber. In the limit that $q \rightarrow 0$, the Hough functions become ordinary (appropriately normalized) associated Legendre functions, while $\lambda\rightarrow l(l+1)$.

We present the inhomogeneous, tidally driven stellar oscillation equations in the traditional approximation in \ap{a:trad}. The principal difference relative to the standard stellar oscillation equations is that terms involving $l(l+1)$ either are approximated to zero, or have $l(l+1)\rightarrow\lambda$. This replacement changes the effective angular wavenumber. E.g., since the primary $\lambda$ for $m=0$ increases with increasing rotation, fast rotation leads to increased damping of $m=0$ g-modes at fixed frequency, as discussed in \se{s:rotres}.

For strong rotation, $|q|>1$, the Hough eigenvalues $\lambda$ can be both positive and negative. The case of $\lambda > 0$ produces rotationally modified traditional g-modes, which evanesce for $\cos^2\theta > 1/q^2$. (Rossby waves or r-modes are also confined near the equator and have a small positive value of $\lambda$.) Instead, for $\lambda<0$, polar modes are produced that propagate near the poles for $\cos^2\theta > 1/q^2$, but evanesce radially from the surface since they have an imaginary Lamb frequency $S_\lambda=\lambda^{1/2} c_s/r$ (as explained further in \fig{f:prop}).

\section{Qualitative discussion of tidal asteroseismology}\label{s:qualres}
It is helpful conceptually to divide the tidal response of a star into two components, the \emph{equilibrium tide} and the \emph{dynamical tide} \citep{zahn75}. Note that in this section we will again use the normal mode formalism described in \se{s:nmodes}, even though our subsequent more detailed modeling of \koi{} uses the inhomogeneous, nonadiabatic formalism introduced in \se{s:nonad}.

\subsection{Equilibrium tide}\label{s:qualeq}
The equilibrium tide is the ``static'' response of a star to a perturbing tidal potential, i.e., the large-scale prolation due to differential gravity from a companion. In terms of lightcurves, the equilibrium tide corresponds to ellipsoidal variability (along with the irradiation component of this effect discussed in \ap{a:reflect}). In the case of an eccentric binary this manifests itself as a large variation in the observed flux from the binary during periastron. \koi's equilibrium tide was successfully modeled in \welsh{}, enabling precise constraints to be placed on various stellar and orbital parameters (\ta{t:wparams}).

In terms of the normal mode formalism developed in \se{s:nmodes}, the equilibrium tide corresponds to the amplitudes from \eq{e:nmodes} tied to large overlaps $\overlap$ and large Hansen coefficients $\hans$; in other words, to pairings of low-$|n|$ modes with low-$|k|$ orbital harmonics. The Lorentzian factor $\lorentz$ is typically $\sim$  1 for the equilibrium tide since it is not a resonant phenomenon.

In practice, however, it is much simpler and more convenient to use other mathematical formalisms to model the equilibrium tide, like taking the zero-frequency stellar response as in \ap{a:eqtide}, or filling Roche potentials as in \welsh{}'s simulations.   We show in \se{s:ellips} that our simple analytical treatment of the equilibrium tide verifies the results from the sophisticated simulation code employed in \welsh.

\subsection{Dynamical tide}\label{s:qualdyn}
The dynamical tide, on the other hand, corresponds to resonantly excited  pulsations with frequencies equal to harmonics of the orbital frequency, $k\omegorb$. \welsh{} observed at least 21 such harmonics (\ta{t:data}).

For a circular orbit, the tidal potential has all its power in the $k=\pm2$ orbital harmonics; in this case the only modes that can be resonantly excited are those with frequencies close to twice the Doppler-shifted orbital frequency: $\omegnl \approx 2|\omegorb - \omegrot|$; this is typically only a single mode. This corresponds to the fact that the Hansen coefficients from \eq{e:hans} become a Kronecker delta at zero eccentricity: $\hansnorm(0) = \delta_{m}^k$. However, for a highly eccentric orbit, the distribution of power in the Hansen coefficients, and hence the stellar response, can be much broader; as a result a wide array of different harmonics can be excited, allowing for a rich pulsation spectrum.

\label{s:pureharm}Mode excitation due to a tidal harmonic $k\omegorb$ is modulated by the Doppler-shifted frequency $\dopp = k\omegorb - m\omegrot$. However, the frequencies at which modes are \emph{observed} to oscillate, viewed from an inertial frame, are indeed pure harmonics of the orbital frequency, $k\omegorb$.\footnote{\citet{welsh11} incorrectly attributed nonharmonic pulsations to rotational splitting; we return to nonharmonic pulsations in \se{s:anom}.} We demonstrate this mathematically in \ap{a:nonad}; intuitively, although a driving frequency experiences a Doppler shift upon switching to a star's corotating frame, the star's response is then Doppler shifted back upon observation from an inertial frame. In general, any time a linear system is driven at a particular frequency, it then also oscillates at that frequency, with its internal structure reflected only in the oscillation's amplitude and phase.

Whether a given mode is excited to a large amplitude is contingent on several conditions---essentially all the terms in \eq{e:nmodes}. First, the overall strength of the tide, and hence the magnitude of observed flux variations, is determined by the tidal factor $\epsilon_l$ from \eq{e:tidefac}. The dominant multipole order is $l=2$, so we have $\epsilon_2 = (M_2/M_1)(R_1/\Dp)^3$, where $\Dp=a(1-e)$ is the binary separation at periastron, and we are focusing our analysis on star 1. For \koi, $\epsilon_2 \simeq 4\times 10^{-3}$ for both stars.

Next, the strength of a mode's resonant temporal coupling to the tidal potential is given by the Lorentzian factor $\lorentz$ in \eq{e:lorentz}. Since this factor is set by how close a mode's frequency is to the nearest orbital harmonic, its effect is intrinsically random. The degree of resonance has an enormous effect on a mode's contribution to the observed flux perturbation, meaning that modeling the dynamical tide amounts on some level to adjusting stellar and system parameters in order to align eigenfrequencies against orbital harmonics so that the array of Lorentzian factors conspire to reproduce observational data.

Moreover, given a single observed pulsation amplitude together with theoretical knowledge of the likely responsible mode, i.e.\ the first four factors in \eq{e:nmodes}, equating theoretical and observed pulsation amplitudes in principle yields direct determination of the mode's eigenfrequency, independently of the degree of resonance. This line of reasoning of course neglects the considerable theoretical uncertainties present, but serves to illustrate tidal asteroseismology's potential to constrain stellar parameters. 

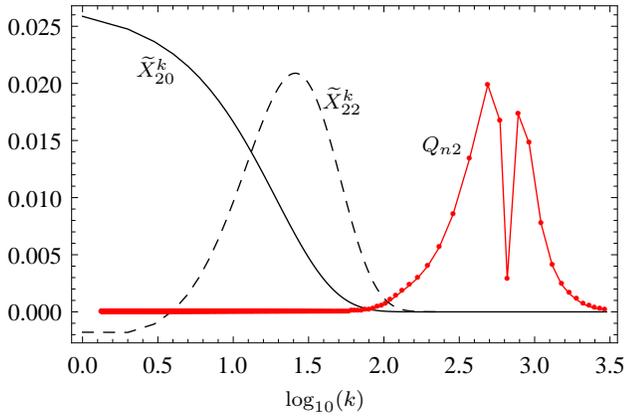
\begin{figure}
  \begin{overpic}{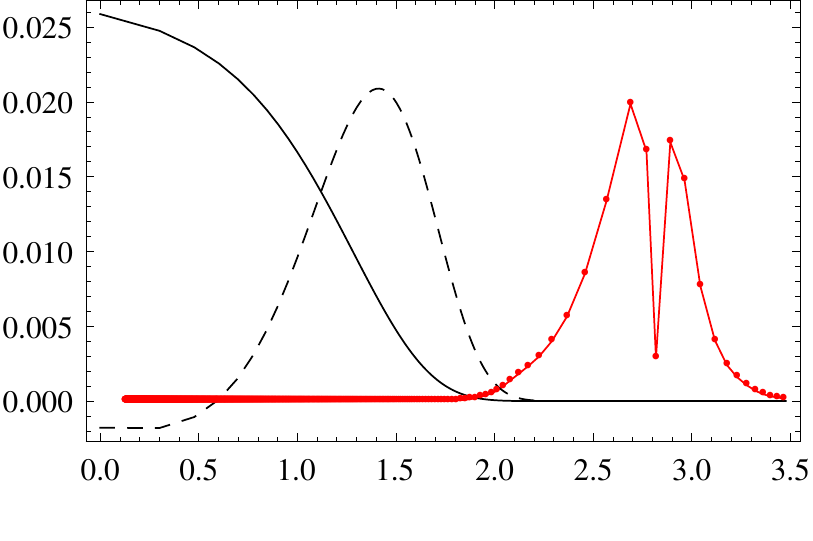}
    \put(45,2){$\log_{10}(k)$}
    \put(21,55){$\widetilde{X}_{20}^k$}
    \put(51,50){$\widetilde{X}_{22}^k$}
    \put(67,43){$Q_{n2}$}
  \end{overpic}
  \caption{The linear overlap integral $Q_{nl}$ and unit-normalized Hansen coefficients $\hansnorm(e)$ as a function of $k=\omega/\omegorb$ for $l=2$ and $m=0,2$ (note the identity $\hansnorm = \widetilde{X}_{l,-m}^{-k}$). Stellar and orbital parameters are fixed to \welsh's mean values for star 1; in particular, $e=0.8342$. Each point on the curve for $Q$ represents a normal mode with frequency $\omegnl \simeq k\omegorb$, thus neglecting any Doppler shift due to rotation. (See \se{s:rotres} for a discussion of the influence of Doppler shifts.) Modes of a given $m$ can be excited near  where the overlap curve intersects the Hansen curve, at $\log_{10}(k) \approx 2.0$ in this plot.}
  \label{f:hansoverlapsep}
\end{figure}

Despite the inherent unpredictability, a lightcurve's Fourier spectrum is still subject to restrictions imposed primarily by the remaining two factors in \eq{e:nmodes}.
These terms, the linear overlap integral $Q_{nl}$ and the unit-normalized Hansen coefficient $\hansnorm(e)$ (respectively equations \ref{e:overlap} and \ref{e:hans}), restrict the range in $k$ over which harmonics can be excited; \fig{f:hansoverlapsep} shows profiles of both. As discussed in \se{s:nmodes}, $Q_{nl}$ peaks for modes with frequencies near the dynamical frequency of the star $\omegdyn$ and falls off as a power law in frequency, whereas $\hansnorm$ peaks for harmonics near $m\omegperi/\omegorb$ and falls off for higher $k$:
\begin{align}
 &&Q_{nl}  &\propto \omegnl^p & \omegnl &\ll \omegdyn&&\\
 &&\hansnorm(e)  &\propto \exp(-k/r) & |k| &\gg m\omegperi/\omegorb.&&
\end{align} 
The power-law index $p$ is 11/6 for g-modes in stars with a convective core and a radiative envelope or vice versa \citep{zahn70}, and for \koi{}'s eccentricity and $l=2$ we find $r \sim 15$.

\begin{figure}
  \begin{overpic}{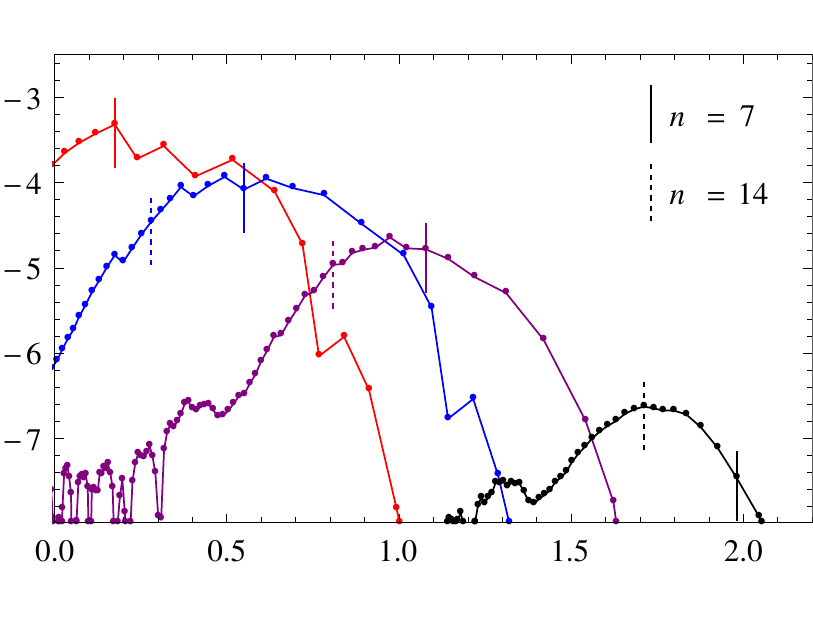}
    \put(45,2){$\log_{10}(k)$}
    \put(1,74){$\log_{10}\left(Q_{n2}^{\phantom{k}} \widetilde{X}_{20}^k(e)\right)$}
    \put(18,61){$e=0.2$}
    \put(40,55){$e=0.4$}
    \put(63,43){$e=0.6$}
    \put(85,27){$e=0.8$}
  \end{overpic}
  \caption{Plot of $Q_{nl} \hansnorm(e)$ as a function of $k=\omega/\omegorb$ for several eccentricities with $l=2$, $m=0$, and all stellar parameters as well as the periastron separation $\Dp$ fixed to \welsh's mean values for star 1 (\ta{t:wparams}). Fixing $\Dp$ fixes the tidal factor $\epsilon_l$ from \eq{e:tidefac} and hence the overall strength of the tide (although the orbital period consequently varies). Each point represents a normal mode; Hansen coefficients are evaluated with $k$ given by the integer nearest to $\omegnl/\omegorb$ for each eigenmode. Solid vertical lines denote the $n=7$ g-mode (the g-mode with 7 radial nodes), while dashed vertical lines are $n=14$. Note that the finer eigenfrequency spacing at small $k$ allows for larger amplitudes, which is not accounted for in this plot; including this effect would shift the curves toward lower $k$.}
  \label{f:hansoverlapprod}
\end{figure}

As a result, modes that can be excited are those with frequencies in the intervening region between the peaks of $Q_{nl}$ and $\hans$, i.e.,
\begin{equation}
 |m|\omegperi < \omegnl < \omegdyn.
\end{equation}
This is a necessary but not sufficient condition; \fig{f:hansoverlapprod} shows the product $Q_{nl}\hans(e)$ at various eccentricities with stellar parameters as well as the periastron distance $\Dp$ fixed to the mean values in \welsh, and hence with fixed tidal parameter $\epsilon_l$ (but consequently allowing the orbital period to vary). Although a chance close resonance can yield a large Lorentzian factor $\lorentz$, excitation of modes far from the peak of $Q_{nl}\hansnorm$ becomes less and less likely, since this quantity falls off sharply, especially towards larger $|k|$.

There are two other constraints on the range of harmonic pulsations that can be excited. First, the eigenmode density for g-modes scales asymptotically as 
\begin{equation}
\left|\frac{dn}{dk}\right| \sim \frac{l}{k^2}\cdot\frac{\omegdyn}{\omegorb}, \end{equation}
which shows that fewer modes exist at higher $k$. This can be seen by the spacing of points (which denote normal modes) in Figures \ref{f:hansoverlapsep} and \ref{f:hansoverlapprod}, as well as by the spacing of peaks in \fig{f:nonad_fine}. This further limits the number of harmonics that can be excited at large $k$, in addition to the exponential decay of the Hansen coefficients discussed earlier, and thus effectively shifts the curves in \fig{f:hansoverlapprod} toward lower $k$.

In addition, the Lorentzian factor $\lorentz$ is attenuated by mode damping $\gamma_{nl}$, which is set by radiative diffusion for high-order g-modes. Damping becomes larger with decreasing g-mode frequency due to increasing wavenumber; an asymptotic scaling is given in \eq{e:gamscale}. Because the Lorentzian response is proportional to $\gamma_{nl}^{-1}$ at perfect resonance, the amplitudes of lower-frequency modes/harmonics are  diminished by increased damping, in addition to the power-law decay of the tidal overlap. This effect is critical for understanding the influence of rotation on lightcurve power spectra, as we investigate in \se{s:rotres}.

\subsection{Pulsation phases}\label{s:phase}
Pulsation phases in eccentric binaries are essential information which should be fully modeled, in addition to the pulsation amplitudes reported in \welsh. For simplicity, we focus on a particular harmonic amplitude $A_{nlmk}$ from equations \eqref{e:summodes} and \eqref{e:nmodes} and assume it results from a close resonance so that $\omegnl\approx\dopp=k\omegorb - m\omegrot$, assuming without loss of generality that $\dopp>0$. We can then evaluate its phase $\psi_{nlmk}$ relative to periastron, modulo $\pi$ (since we are temporarily ignoring the real part of the amplitude, which could introduce a minus sign), as
\begin{equation}\begin{split}\label{e:Aphase}
 \psi_{nlmk}&=\arg(A_{nlmk})\\
  & = \pi/2-\arctan\left(\frac{\delta\omega(\omegnl+\dopp)}{2\gamma_{nl}\dopp}\right)\mod \pi\\
  & \approx \pi/2-\arctan(\delta\omega/\gamma_{nl})\mod \pi,
\end{split}\end{equation}
where $\delta\omega=\omegnl-\dopp$ is the detuning frequency.

For a near-perfect resonance, where $\delta\omega \lesssim \gamnl$, $\psi_{nlmk}$ approaches $\pi/2$ (modulo $\pi$). However, if eigenmode damping rates are much smaller than the orbital frequency, then this intrinsic phase should instead be near 0. This is the case for \koi{}, where $\omegorb/\gamnl>10^{3}$ for modes of interest. Indeed, theoretically modeling the largest-amplitude 90th and 91st harmonics of \koi{} assuming they are $m=0$ modes requires only $|\delta\omega|/\gamnl\sim20$, so that even these phases should be within $\sim 1\%$ of zero (modulo $\pi$).

The phase of the corresponding observed harmonic flux perturbation can be obtained from \eq{e:Aphase} by further including the phase of the spherical harmonic factor in the disk-averaging formula, \eq{e:fluxvar}:
\begin{equation}\label{e:dJphase}
 \arg(\delta J_k/J) = \psi_{nlmk} + m\phi_o.
\end{equation}
Summing over the complex conjugate pair, the observed time dependence is then $\cos[k\omegorb t - (\psi_{nlmk} + m\phi_o)]$, where $t=0$ corresponds to periastron. However, the sign of $k$ in this formalism is unknown; equivalently, whether the pulsation is prograde or retrograde (\ap{a:torque}) cannot be determined in this way. Thus if the observed pulsation's (cosine) phase is $\delta$, the comparison to make is
\begin{equation}\label{e:obsphase}
 \delta = \pm (\psi_{nlmk} + m\phi_o) \mod \pi.
\end{equation} 
Nonetheless, since we have argued that $\psi_{nlmk}\approx0$, this becomes
\begin{equation}\label{e:obsphasesimp}
 \delta \approx \pm m\phi_o \mod \pi.
\end{equation}

Given determination of $\phi_o$ (related to the argument of periastron $\omega$ by \eqp{e:obsang}) by modeling of RV data or ellipsoidal variation, the phase of a resonant harmonic thus directly gives the mode's value of $|m|$ (which is very likely 0 or $2$ for tidally excited modes, since $l=2$ dominates). For \koi, phase information on harmonics 90 and 91 would thus determine whether they result from  resonance locks, as discussed in the next section, or are simply chance resonances. Furthermore, knowing $|m|$ allows $m\phi_o$ to be removed from \eq{e:obsphase}, yielding the pulsation's damping-to-detuning ratio.

However, the preceding treatment is only valid if the eigenfunction itself has a small phase: although eigenfunctions are purely real for adiabatic normal modes, local phases are introduced in a fully nonadiabatic calculation, as in \se{s:nonad}. Thus equations \eqref{e:dJphase} -- \eqref{e:obsphasesimp} are only applicable in the standing wave regime, where the imaginary part of the flux perturbation is small relative to the real part. In the traveling wave regime, the local wave phase near the surface becomes significant, and can overwhelm the contribution from global damping; see \se{s:nonad}. For \koi, this corresponds to $|k|$ below $\sim 30$, although this depends on the rotation rate (\se{s:rotres}).

\section{Rotational synchronization in \koi}\label{s:sync}
Here we will discuss \emph{a priori} theoretical expectations for \koi's stars' rotation. Later, in \se{s:rotres}, we will compare the results derived here with constraints imposed by the observed pulsation spectrum.

\subsection{Pseudosynchronization}\label{s:pseudo}
In binary systems, the influence of tides causes each component of the binary to eventually synchronize its rotational and orbital motions, just as with Earth's moon. Tides also circularize orbits, sending $e\rightarrow 0$, but the circularization timescale $\tc$ is much greater than the synchronization timescale $\ts$; their ratio is roughly given by the ratio of orbital to rotational angular momenta:
\begin{equation}\begin{split}
\frac{\tc}{\ts} &\sim \frac{L_\mr{orb}}{L_\mr{*}} =\frac{\mu a^2}{I_*}\cdot\frac{\omegorb}{\omegrot}\cdot\sqrt{1-e^2}\\
&\sim \left(\frac{a}{R}\right)^2\left(\frac{M_* R_*^2}{I_*}\right)(1-e)^2,\\
\end{split}\end{equation}
where $I_*$ is the stellar moment of inertia, $\mu$ is the reduced mass, and we have assumed for simplicity that the stars rotate at the periastron frequency $\omegperi$ (\se{s:prelim}). For \koi, this ratio is $\sim 10^3$.

Due to the disparity of these timescales, a star in an eccentric binary will first synchronize to a \emph{pseudosynchronous period} $\Pps$, defined as a rotation period such that no average tidal torque is exerted on either star over a sufficiently long timescale. If only the torque due to the equilibrium tide is used, and thus eigenmode resonances are neglected, then only one unique pseudosynchronous period exists, $\Ppse$, as derived in \citet{hut81} and employed in \welsh. Its value for \koi{} is (\eqp{e:hutpseudo})
\begin{displaymath}
 \Ppse = (2.53\pm0.01)\ \mr{days}.
\end{displaymath}

\begin{figure}
\centering
\begin{tabular}{@{}r@{\ }c}
 & $\quad \Prot$ (day)\rule[-5pt]{0pt}{0pt} \\
 \rotatebox{90}{\rule{37pt}{0pt}$\log_{10}\big[\,\tau/(GM^2/R)\,\big]$} & \begin{overpic}{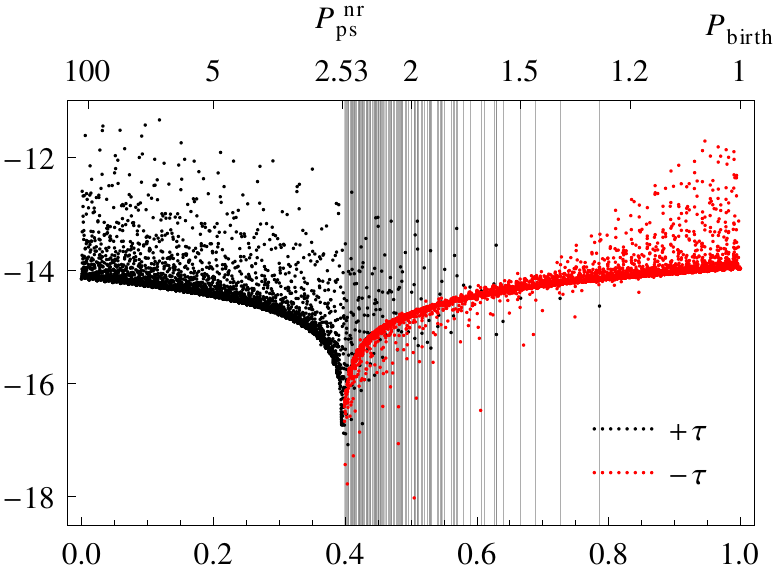}\end{overpic}\\
 & $\quad \frot$ ($1/\mr{day}$)\rule[-10pt]{0pt}{0pt}\\
 & $\quad \Prot$ (day)\rule[-5pt]{0pt}{0pt} \\
 \rotatebox{90}{\rule{43pt}{0pt}$\tau/(GM^2/R)\times10^{14}$} & \begin{overpic}{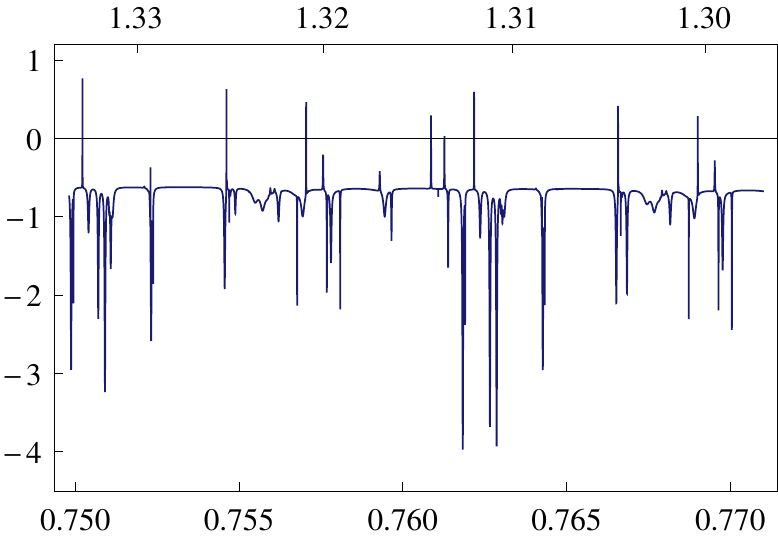}\end{overpic}\\
 & $\quad \frot$ ($1/\mr{day}$)
\end{tabular}
  \caption{Plot of the secular tidal synchronization torque $\tau$ as a function of the rotational frequency $\frot = 1/\Prot$ for star 1, using \welsh's parameters (\ta{t:wparams}). {\bf Top panel:} Black indicates a positive torque (meaning increasing stellar spin), while red indicates a negative torque (decreasing stellar spin). The overall profile of $\tau$ goes to zero at Hut's value for the nonresonant pseudosynchronous period, $\Ppse=2.53$ days for \koi, but eigenmode resonances create many additional zeroes, displayed as light gray vertical lines. (Determination of zeroes in this plot is limited by its grid resolution. Many more exist; see bottom panel.) {\bf Bottom panel:} Zoom in, showing how narrow resonance spikes can cause the otherwise negative torque to become zero. The torque pattern roughly repeats at half the orbital frequency, $f_\mr{orb}/2=0.012\ \text{day}^{-1}$, due to the fact that eigenmodes resonate with Doppler-shifted driving frequencies $\dopp=k\omegorb-m\omegrot$, with $m=\pm2$ and $k\in\mathbb{Z}$.}
  \label{f:torque}
\end{figure}

Inclusion of eigenmode resonances, however, complicates the situation. \fig{f:torque} shows the secular tidal torque (averaged over one rotation period) for star 1 of \koi{} plotted as a function of rotation frequency/period including contributions from both the equilibrium and dynamical tide. Although the general torque profile tends to zero at $\Ppse$, numerous other roots exist (displayed as vertical lines), where the torque due to a single resonantly excited eigenmode of the dynamical tide cancels against that due to the equilibrium tide. To produce this plot, we directly evaluated the secular tidal torque (\ap{a:torque}) using an expansion over the quadrupolar adiabatic normal modes of a MESA stellar model \citep{paxton11} with parameters set by \welsh's mean values for star 1 of \koi{} (\ta{t:wparams}). In our calculation we include both radiative (\se{s:nmodes}) and turbulent convective damping \citep{willems10}, but neglect rotational modification of the eigenmodes.

Next, of the many zeroes of the secular torque available, which are applicable? Continuing with the assumption that \koi's stars were born with rotation periods of $\Pbirth\sim 1\ \mr{day}$ (\se{s:irot}), with the same orientation as the orbital motion, one might naively posit that the first zero encountered by each star should constitute a pseudosynchronous period---it is an ostensibly stable spin state since small changes to either the stellar eigenmodes (via stellar evolution) or the orbital parameters (via circularization and orbital decay) induce a restoring torque. This is the basic idea behind a resonance lock \citep{witte99}.

However, this conception of resonance locking neglects two important factors. First, although the dynamical and equilibrium tidal torques may cancel, their energy deposition rates do not (in general); see \ap{a:torque}. Thus during a resonance lock the orbital frequency must continue to evolve, allowing other modes to come into resonance, potentially capable of breaking the lock. Second, as shown by \citet{fuller11}, it is necessary that the orbital frequency not evolve so quickly that the restoring torque mentioned earlier be insufficient to maintain the resonance lock. This restricts the range of modes capable of resonantly locking, introducing an upper bound on their inertial-frame frequencies and hence their orbital harmonic numbers (values of $k$ in our notation).

Consequently, pseudosynchronization is in reality a complicated and dynamical process, consisting of a chain of resonance locks persisting until eventually $e\rightarrow 0$ and $\Prot=\Porb$. Such resonance lock chains were studied in much greater detail by \citet{witte99} for eccentric binaries broadly similar to \koi. As a result of the inherent complexity, a full simulation of \koi's orbital and rotational evolution is required in order to address the phenomenon of resonance locking and to derive theoretical predictions for the stars' spins. To perform such simulations, we again expanded the secular tidal torque and energy deposition rate over normal modes (detailed in \ap{a:torque}) using two MESA stellar models consistent with \welsh's mean parameters for \koi's two stars.  We then numerically integrated the orbital evolution equations \citep{witte99} assuming rigid-body rotation. We did not include the Coriolis force, nor did we address whether the eigenmode amplitudes required to produce the various resonance locks that arise are stable to nonlinear processes (\se{s:anom}).

Our simulations indicate that both stars should have reached pseudosynchronization states with rotation periods of $\Pps\sim1.8$ days; we discuss the synchronization timescale in more detail in \se{s:synctime}. These periods are $\sim 30\%$ faster than Hut's value of $\Ppse=2.5$ days. The pseudosynchronization mechanism that operates is stochastic in nature, in which the dynamical tide's prograde resonance locks balance the equilibrium tide in a temporally averaged sense. This result appears independently of the initial rotation rates used; in other words, it is an attractor.

As described above, when a star is locked in resonance, it is the torque from a single highly resonant eigenmode that acts to oppose the equilibrium tide's nonzero torque. Such a high-amplitude mode should be easily observable. At first glance, this line of reasoning seems to provide a natural explanation for the presence of the large-amplitude 90th and 91st observed harmonics in \koi{} (F1 and F2 from \ta{t:data}), namely that each is the photometric signature of the highly resonant eigenmode that produces a resonance lock for its respective star. There are several problems with this idea, however, which we elucidate in \se{s:psmode}.

\subsection{Synchronization timescale}\label{s:synctime}
Where between the stars' putative birth rotation periods, $\Pbirth=1.0$ day, and the pseudosynchronous period from our simulations, $\Ppsp\sim1.8$ days, do we \emph{a priori} expect the rotation periods of \koi's stars to fall? To this end, we can roughly estimate the synchronization timescale $\ts$ by integrating $I\dot{\Omega}_* = \tau(\Omega_*)$ to find
\begin{equation}
\ts \sim I\int_{\omegbirth}^{\omegpsp}\frac{d\,\Omega_*}{\tau(\Omega_*)},
\end{equation}
where $I$ is the stellar moment of inertia, $\omegps = 2\pi/\Pps$, and the tidal torque $\tau$ can as before be calculated as a function of the spin frequency $\Omega_*$ using an expansion over normal modes (\ap{a:torque}).

Using this approximation, we find $\ts \sim 80\ \mr{Myr}$, which is less than the inferred system age of $t_\mr{age}\sim 200\ \mr{Myr}$. This is consistent with our orbital evolution simulations (\se{s:pseudo}). Although stellar evolution was ignored in this calculation, a rough estimate of its effect can be made using only the fact that $\ts$ scales as $R^{-3}$ (since the torque scales as $R^5$ while the moment of inertia scales as $R^2$). Given that both stars had $10\%$ smaller radii at ZAMS (indicated by our modeling), this would lead to only at most a $3\times10\% = 30\%$ increase in $\ts$. Furthermore, both stars had much larger radii before reaching the main sequence, which would imply an even shorter synchronization time. Lastly, an important effect that arises when rotation is fully included is the existence of retrograde r-modes, which would also enhance the rate of stellar spindown \citep{witte99}. Thus the inequality
\begin{displaymath}
 t_\mr{age} > \ts
\end{displaymath}
seems to be well satisfied, and we expect that both stars' rotation periods should be close to the value of $\Ppsp\sim 1.8$ days from \se{s:pseudo}.

\section{Results}
\label{s:results}
\subsection{Ellipsoidal variation}\label{s:ellips}

\begin{figure*}
   \begin{tabular}{@{}r@{}c@{\rule{2pt}{0pt}}c@{\rule{2pt}{0pt}}c@{\rule{2pt}{0pt}}}
    & \rule{15pt}{0pt}(a) & (b) & (c)\\
    & \rule{15pt}{0pt}\koi: $i=5.5\deg$, $\omega=36\deg$ & $i=90\deg$, $\omega=80\deg$ & $i=90\deg$, $\omega=20\deg$\\
    \rotatebox{90}{\rule{50pt}{0pt}$(\delta J/J)\times 10^3$} & \includegraphics{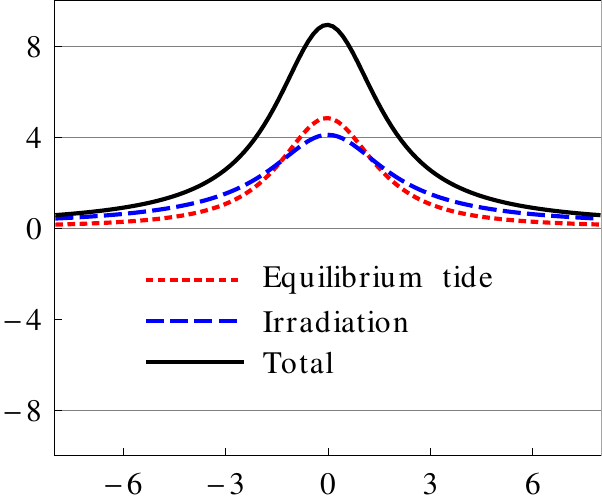} & \includegraphics{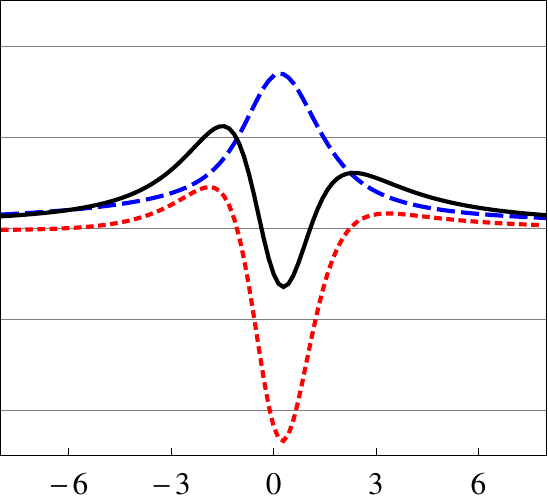} & \includegraphics{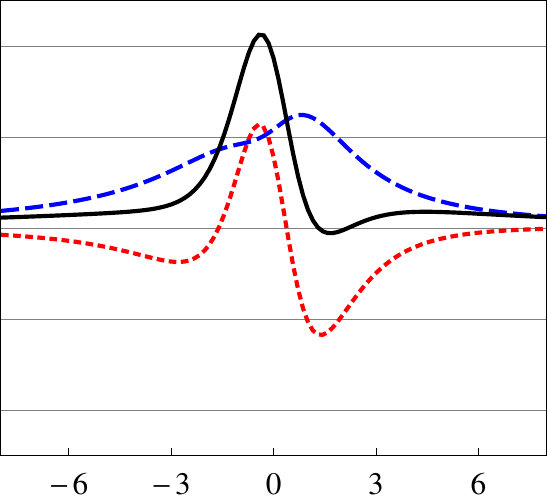}
   \end{tabular}
  \rule{20pt}{0pt}mean anomaly (degrees)
  \caption{Simple analytic model of ellipsoidal variability detailed in \ap{a:ellips}, including both the equilibrium tide (red dotted lines) and ``reflection''/irradiation (blue dashed lines) components of the lightcurve. We used the best-fit parameters from \welsh{} in all three panels (\ta{t:wparams}), except that we show two examples of edge-on orientations (ignoring the possibility of eclipses) in (b) \& (c), effectively presenting \koi's lightcurve as it would be observed from different angles. (We used $\omega=(80\deg,20\deg)$ for (b, c) in order to demonstrate the asymmetric lightcurves possible depending on the binary's orientation.) Panel (a) reproduces \welsh's modeling and the data for \koi{} to $\sim$  20\%. Our analytic model is easily applicable to many other systems. In \se{s:rotres} we show that the dynamical tidal response, ignored here, may be larger than that due to the equilibrium tide for edge-on systems.}
  \label{f:ellips}
\end{figure*}

\fig{f:ellips}.a shows our simple model of \koi's ellipsoidal variation; we adopted the best-fit parameters from \welsh's modeling (\ta{t:wparams}) to produce our lightcurve. Our irradiation (\ap{a:reflect}, blue dashed line) and equilibrium tide (\ap{a:eqtide}, red dotted line) models are larger than \welsh's results by 24\% and 14\% respectively.   The shapes of both curves are, however, essentially indistinguishable from \welsh's much more detailed calculations.

We attribute the small difference between our results and those of \welsh{} to our simple model of the bandpass correction (\eqp{e:bandpass}) which ignores bandpass variations due to limb darkening. Such details could easily be incorporated into our analytical formalism, however, by introducing a wavelength-dependent limb darkening function $h_\lambda(\mu)$ in the disk integrals in equations \eqref{e:diskb} and \eqref{e:diskc} \citep{robinson82}.   We thus believe that the models provided in \ap{a:ellips} should be quite useful for modeling other systems like \koi, due in particular to their analytic simplicity.

We also show in \fig{f:ellips}.b \& c what \koi{}'s equilibrium tide and irradiation would look like for two edge-on orientations, demonstrating the more complicated, asymmetric lightcurve morphologies possible in eccentric binaries (see also the earlier work by \citealt{kumar95}). Future searches for eccentric binaries  using Kepler and other telescopes with high-precision photometry should allow for the wide range of lightcurve shapes shown in \fig{f:ellips}. We note, however, that that the dynamical tidal response, ignored in this section, may be larger than that due to the equilibrium tide for edge-on systems, as we show in \se{s:rotres}.

\subsection{Nonadiabatic inhomogeneous method}\label{s:nonad}
Thus far our theoretical results have primarily utilized the tidally forced adiabatic normal mode formalism. Although this framework provides excellent intuition for the key physics in eccentric binaries, it is insufficient for  producing detailed theoretical lightcurves, since this necessitates tracking a star's tidal response all the way to the photosphere where nonadiabatic effects are critical.   To account for this,  we employ the nonadiabatic inhomogeneous formalism originally used by \citet{pfahl08} (\ap{a:zeroth}), which we have extended to account for rotation in the traditional approximation (\ap{a:trad}).

\begin{figure*}
  \begin{overpic}{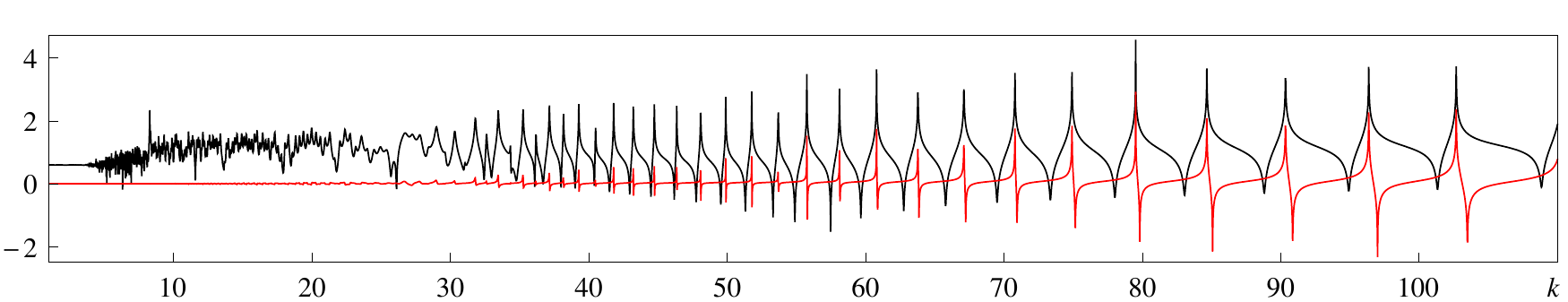}
    \put(.5,18.3){$\log_{10}$}
    \put(11,5){$|\xi_r/R|_\mr{phot}$}
    \put(10,12.2){$|\Delta F/F|_{\mr{phot}}$}
  \end{overpic}
  \caption{Amplitude of both the radial displacement $\xi_r$ and Lagrangian flux perturbation $\Delta F$ evaluated at the photosphere as a function of $k=\omega/\omegorb$ for a MESA  model of star 1, using \welsh's best-fit parameters (\ta{t:wparams}). As a result of its larger amplitude, the Lagrangian flux perturbation has a much larger effect than the surface displacement on observed flux variations. Rotation is not included in this calculation. The radial dependence of the tidal potential is set to $U=-(GM_1/R_1)(r/R_1)^{l}$ with $l=2$, which determines the vertical scale. Normal mode eigenfrequencies correspond to peaks in the curves. See the text for a discussion of the different regimes present in the stellar response for different forcing frequencies.}
  \label{f:nonad_fine}
\end{figure*}

Rather than decompose the response of the star into normal modes, the inhomogeneous method directly solves for the full linear response of the star to an external tidal force produced by a companion at a given forcing frequency. Given a stellar model, an orbital period, a set of orbital harmonics to act as driving frequencies, and a rigid-body rotation period, we solve the numerical problem described in \ap{a:trad} for each star. This determines the various physical perturbation variables of the star as a function of radius, such as the radial displacement and the flux perturbation. For stars of interest we can safely ignore perturbations to the convective flux, so the only nonadiabatic effect is that produced by radiative diffusion.

\fig{f:nonad_fine} shows the surface radial displacement $\xi_r$ and Lagrangian emitted flux perturbation $\Delta F$ computed on a fine frequency grid, temporarily ignoring rotation; normal mode frequencies correspond to the resonant peaks in these curves. The surface radial displacement should approach its equilibrium tide value as the driving frequency tends to zero.   Quantitatively, we find that this is true for orbital harmonics $k\lesssim 30$; note that in the units employed in \fig{f:nonad_fine}, this equilibrium tide value for $\xi_r$ is $(\xi_r/R)_\mr{phot}^{\mr{eq}}= 1$ (\ap{a:eqtide}).

The surface flux perturbation shown in \fig{f:nonad_fine}, on the other hand, more clearly demonstrates the three qualitatively different regimes possible at the surface. First, the weakly damped standing wave regime, $k\gtrsim 30$, is characterized by strong eigenmode resonances and all perturbation variables having small imaginary parts. In \fig{f:prop}, this corresponds to the outer turning point, where the mode frequency intersects the Lamb frequency, lying inside the point where the mode frequency becomes comparable to the thermal frequency, so that the mode becomes evanescent before it becomes strongly nonadiabatic.

Next, the traveling wave regime, $5\lesssim k\lesssim 30$, arises when modes instead propagate beyond where the mode and thermal frequencies become comparable, leading to rapid radiative diffusion near the surface.  In the traveling wave limit, resonances become severely attenuated as waves are increasingly unable to reflect at the surface, and all perturbation variables have comparable real and imaginary parts (not including their equilibrium tide values).

Lastly, just as with the radial displacement, the flux perturbation also asymptotes to its overdamped equilibrium tide/von Zeipel value of $(\Delta F/F)_{\mr{phot}}^\mr{eq,vZ}= -(l+2)(\xi_r/R)_\mr{phot}^{\mr{eq}}$ (\ap{a:eqtide}) in the low-frequency limit, which is $|\Delta F/F|_{\mr{phot}}^\mr{eq,vZ}=|-l-2|=4$ in \fig{f:nonad_fine}'s units. Quantitatively, however, this only occurs for $k\lesssim 5$. At first glance, this suggests that the equilibrium tide modeling of \koi{} in \welsh{} and \fig{f:ellips} is invalid, since the equilibrium tide in \koi{} has orbital power out to at least $k\sim 30$ (as can be seen e.g.\ in the plot of the Hansen coefficients for \koi's eccentricity in \fig{f:hansoverlapsep}).

Fortunately, as we describe in the next section, including rotation with a face-on inclination effectively stretches the graph in \fig{f:nonad_fine} towards higher $k$. E.g., for $\Prot=2.0$ days, we find the equilibrium tide/von Zeipel approximation to hold for $k\lesssim 30$, justifying the simplifications used in \welsh{} and \ap{a:eqtide}, although this may not apply for edge-on systems.

\subsection{Effect of rotation on the dynamical tidal response}
\label{s:rotres}
\begin{figure*}
\flushleft
(a) Face on: $i=5.5\deg$, $\omega=36\deg$ (\koi's orientation)\vspace{.1cm}

   \begin{tabular}{@{}r@{\rule{2pt}{0pt}}c@{\rule{2pt}{0pt}}c@{\rule{2pt}{0pt}}c@{\rule{2pt}{0pt}}c}
      $\Big[\ \Prot/\mr{day}=$\rule{10pt}{0pt} 1.0 \rule{38pt}{0pt}& 1.5 & 2.0 & \phantom{\rule{35pt}{0pt}$\Big]$}$\infty$\rule{35pt}{0pt}$\Big]$ \rule[-9pt]{0pt}{0pt}& \koi\\
    \rotatebox{90}{\rule{15pt}{0pt}$|\delta J_k/J|\times 10^4$} \includegraphics{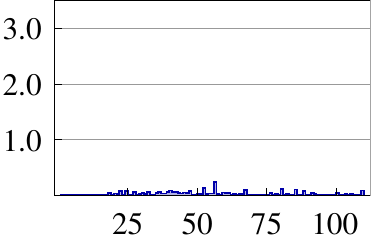} & \includegraphics{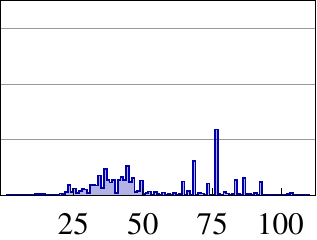} & \includegraphics{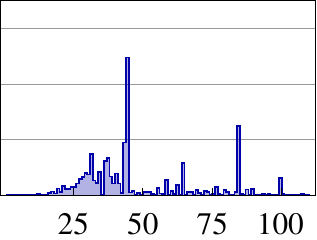} & \includegraphics{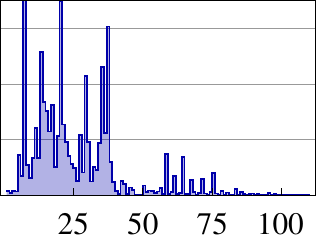} & \begin{overpic}{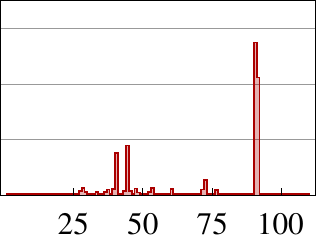}\put(102,2.5){$k$}\end{overpic}
   \end{tabular}\vspace{.1cm}

(b) Edge on: $i=90\deg$, $\omega=36\deg$\vspace{.1cm}

   \begin{tabular}{@{}r@{\rule{2pt}{0pt}}c@{\rule{2pt}{0pt}}c@{\rule{2pt}{0pt}}c@{\rule{2pt}{0pt}}c}
      $\Big[\ \Prot/\mr{day}=$\rule{10pt}{0pt} 1.0 \rule{38pt}{0pt}& 1.5 & 2.0 & \phantom{\rule{35pt}{0pt}$\Big]$}$\infty$\rule{35pt}{0pt}$\Big]$ \rule[-9pt]{0pt}{0pt}& Equilibrium tide\\
    \rotatebox{90}{\rule{15pt}{0pt}$|\delta J_k/J|\times 10^4$} \includegraphics{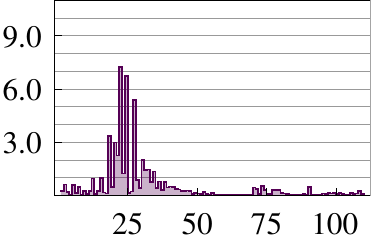} & \includegraphics{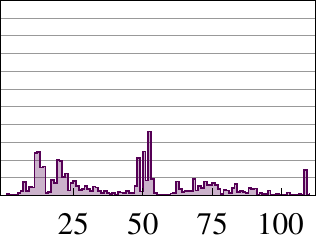} & \includegraphics{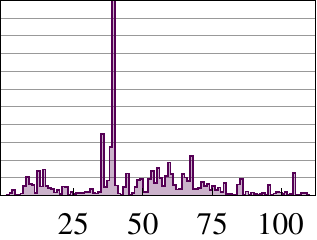} & \includegraphics{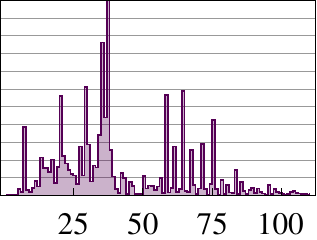} &  \begin{overpic}{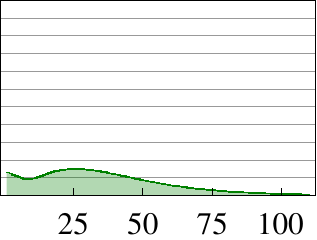}\put(102,2.5){$k$}\end{overpic}
   \end{tabular}
  \caption{Influence of rotation on the lightcurve temporal power spectra of eccentric binaries. The eight leftmost plots show theoretical power spectra using fiducial stellar models and orbital parameters consistent with \welsh's mean values (\ta{t:wparams}), each as a function of orbital harmonic $k$. The top row uses \koi's known face-on inclination of $i=5.5\deg$; the bottom row instead uses an edge-on inclination of $i=90\deg$, showing how a system like \koi{} would appear if viewed edge on. The vertical axis has different scales in the two rows. The effects of rotation on the stellar pulsations are included using the traditional approximation (\se{s:rot}). The equilibrium tide (\ap{a:eqtide}) has been subtracted in order to focus on resonant effects. We have not included negative-$\lambda$ modes, as discussed in the text. (Note that parameters here have not been optimized to reproduce \koi's lightcurve; see \fig{f:results} for such a model.) The four leftmost columns show four different rigid-body rotation periods. {\bf (a)} For a perfectly face-on orientation, only $m=0$ modes can be observed (\se{s:obsflux}); however, larger rotation rates lead to equatorial compression of lower-frequency $m=0$ g-modes, which increases the effective $l$ and hence enhances the dissipation, leading to attenuated amplitudes. The rightmost panel shows the data for \koi{}, which is qualitatively most consistent with shorter rotation periods of 1.0 -- 2.0 days, comparable to the pseudosynchronous  period of $\sim$  1.8 days calculated in \se{s:sync}. {\bf (b)} For edge-on systems, a similar argument regarding rotational suppression of mode amplitudes applies, but instead near the Doppler-shifted harmonic $k = 2\Porb/\Prot$, rather than for $k$ near zero for the $m=0$ modes observable face on (see text for details). Comparison of the four left panels to the rightmost panel, which shows our simple analytical equilibrium tide model's harmonic decomposition (\ap{a:eqtide}), demonstrates that the dynamical tide can dominate the lightcurve in edge-on systems.}
   \label{f:rotres}
\end{figure*}
The most important effects of rotation in the context of tidal asteroseismology can be seen in \fig{f:rotres}. Here we show the predicted flux perturbation for \koi{} as a function of orbital harmonic $k$ for four different rotation periods, having subtracted the equilibrium tide (\ap{a:eqtide}) to focus on resonant effects. \footnote{We assume that rotation is in the same sense as orbital motion throughout this section.} In \fig{f:rotres}.a we use \koi's face-on inclination of $i=5.5\deg$, while in \fig{f:rotres}.b we use an inclination of $i=90\deg$ to illustrate how a system like \koi{} would appear if seen edge on; all other parameters are fixed to those from \welsh's modeling (and are thus not intended to quantitatively reproduce the data; see \fig{f:results} for an optimized model).  The details of which specific higher harmonics have the most power vary as rotation changes mode eigenfrequencies, moving eigenmodes into and out of resonance. Nonetheless, several qualitative features can be observed.

For \koi's actual face-on orientation, as in \fig{f:rotres}.a, rotation tends to suppress power in lower harmonics. This can be understood as follows. Primarily $m=0$ modes are observable face on (\se{s:obsflux}). At fixed driving frequency $\sigma$, as the stellar rotation frequency $\omegrot$, and hence the Coriolis parameter $q=2\omegrot/\sigma$, increases in magnitude, $m=0$ g-modes become progressively confined to the stellar equator (\se{s:rot}). As a result, these rotationally modified modes angularly couple more weakly to the tidal potential, diminishing their intrinsic amplitudes. Moreover, equatorial compression also corresponds to an increase in the effective multipole $l$, where $l\sim\sqrt{\lambda}$, and $\lambda$ is a Hough eigenvalue from \se{s:rot} (e.g., Fig.~2 of \citealt{bildsten96}). Consequently, since g-modes asymptotically satisfy \eq{e:gfreq}, the number of radial nodes $n$ must increase commensurately. Larger $n$ increases the radial wavenumber, which enhances the damping rate, further suppressing the resonant response of the modes and hence their contribution to the observed flux variation.   This effectively corresponds to extending the highly damped traveling wave regime toward higher $k$ in \fig{f:nonad_fine}.

As described in \se{s:rot}, when the magnitude of the Coriolis parameter becomes greater than unity, a new branch of eigenmodes develops with negative Hough eigenvalues, $\lambda < 0$. These modes are confined to the stellar poles rather than the equator \citep{lindzen66}. They also have an imaginary Lamb frequency, so that they are radially evanescent (explained further in \fig{f:prop}), and couple weakly to the tidal potential. We found negative-$\lambda$ modes to produce only a small contribution to the stellar response, which increased with increasing rotation rate but which was roughly constant as a function of forcing frequency, thus mimicking the equilibrium tide. The role of these modes in the context of tidal asteroseismology should be investigated further, but for now we have neglected them in \fig{f:rotres}.

For edge-on orbits, as in \fig{f:rotres}.b, the situation is more complicated, and there are high-amplitude pulsations observable at all rotation periods. First, $m=0$ modes very weakly affect edge-on lightcurves, since their Hansen coefficients (which peak at $k=0$) do not intersect with the linear overlap integrals as strongly as for $m=2$ modes (explained further in \se{s:qualdyn} and shown in \fig{f:hansoverlapsep}). Similarly, modes with $m=-2$ have Hansen coefficients which peak near $-2\omegperi/\omegorb$ and are very small for $k\ge0$.\footnote{It is sufficient to consider only nonnegative $k$, i.e.\ to use a unimodal Fourier series, since the Fourier coefficient of orbital harmonic $k$ must be the complex conjugate of that for  $-k$, since the lightcurve is real valued.} Thus regardless of rotation, only $m=+2$ modes make significant lightcurve contributions.

Within the $m=+2$ modes, there are two regimes to consider: prograde modes excited by harmonics $k>2\omegrot/\omegorb$ and retrograde modes with $k<2\omegrot/\omegorb$ (see \ap{a:torque}). Prograde modes at a given corotating frame frequency are Doppler shifted toward large $k$, whereas the Hansen coefficients peak near $2\omegperi/\omegorb$, so their contribution to lightcurves is marginalized for fast rotation.

Retrograde, $m=+2$ g-modes with small corotating-frame frequencies $\dopp=k\omegorb-m\omegrot$ are subject to the same effect described earlier in the face-on case: they are suppressed by fast rotation due to weaker angular tidal coupling and stronger damping. The difference, however, is that although small driving frequencies are equivalent to small values of $k$ for $m=0$ modes, the Doppler shift experienced by $m=2$ modes means that rotational suppression instead occurs for $k\sim2\omegrot/\omegorb$, which is $84\times(\text{day}/\Prot)$ for \koi's orbital period of 42 days. \fig{f:rotres}.b demonstrates this, where e.g.\ little power can be observed near $k=84$ for $\Prot=1$ day.

Furthermore, rotational suppression does \emph{not} act on low-$k$ harmonics in edge-on systems, as \fig{f:rotres}.b also shows. Indeed, since fast rotation Doppler shifts lower-order retrograde modes---which radially couple more strongly to the tidal potential---toward values of $k$ nearer to the Hansen peak of $\sim2\omegperi/\omegorb$, the power in lower harmonics can even be \emph{enhanced} by sufficiently fast rotation rates.

The rightmost panel of \fig{f:rotres}.b shows the harmonic decomposition of our simple equilibrium tide model for an edge-on orientation, not including irradiation (\se{s:ellips}; \ap{a:eqtide}). Comparing this plot to the left four panels shows in particular that, in edge-on orbits, the dynamical tide is not rotationally suppressed for harmonics where the equilibrium tide has large amplitudes, unlike for face-on orientations. Thus the ellipsoidal variation of edge-on systems may be buried beneath the dynamical tidal response. This implies that full dynamical modeling may be necessary to constrain system parameters for edge-on binaries, and that care must be taken in searches for eccentric binaries, since it cannot be assumed that their lightcurves will be dominated by ellipsoidal modulations.

\subsection{Lightcurve power spectrum modeling}\label{s:model}
We performed preliminary quantitative modeling of the pulsation data in \ta{t:data}. As noted before, tidally driven pulsations should have frequencies which are pure harmonics of the orbital frequency, $\omega = k\omegorb$ for $k\in\mathbb{Z}$. Although most of the pulsations \welsh{} report are of this form, some clearly are not, and are as such unaccounted for in linear perturbation theory. Hence we  only attempted to model pulsations within $0.03$ in $k$ of a harmonic (set arbitrarily); this limited our sample to 21 harmonics, as shown in \ta{t:data}. We provide an explanation for the nonharmonic pulsations in \se{s:anom}.

There are eight primary parameters entering into our modeling of the remaining observed harmonics: stellar masses $M_{1,2}$, radii $R_{1,2}$, ZAMS metallicities $Z_{1,2}$, and rigid-body rotation periods $P_{1,2}$. To explore a range of stellar parameters, we used the stellar evolution code MESA \citep{paxton11} to create two large sets of stellar models, one for each star, with ranges in $M$ and $R$ determined by \welsh's constraints (\ta{t:wparams}). We set both stars' metallicities to 0.04. The other two parameters, $P_1$ and $P_2$, were treated within our nonadiabatic code using the traditional approximation.  We set $P_1 = P_2 =  1.5$ days, comparable to the expected pseudosynchronous rotation period (\se{s:pseudo}) and qualitatively consistent with the small-amplitude flux perturbations of lower harmonics seen in \koi{} (\se{s:rotres}). We fixed all of the orbital parameters to the mean values given in \welsh{}.

As discussed in \se{s:psmode}, it is possible that the 90th and 91st harmonics observed in \koi{} are $m=\pm2$ modes responsible for resonance locks, and are thus in states of nearly perfect resonance.  Indeed, even if they are $m=0$ chance resonances, which are $\sim$  200 times easier to observe with \koi's face-on orbital inclination than $m=\pm 2$ modes (\se{s:obsflux}), we find that a detuning of $|\delta\omega/\omegorb| \sim 10^{-2}$ is required to reproduce the amplitude of either harmonic, where $\delta\omega$ is the difference between the eigenmode and driving frequencies (with $\delta\omega=0$ representing a perfect resonance).

Such a close resonance represents a precise eigenfrequency measurement, and should place stringent constraints on stellar parameters. However, this degree of resonance is also very difficult to capture in a grid of stellar models because even changes in (say) mass of $\Delta M/M \sim 10^{-4}$ can alter the mode frequencies enough to significantly change the degree of resonance; future alternative modeling approaches may obviate this difficulty (\se{s:conc}).  A second problem with trying to directly model the 90th and 91st harmonics is that the amplitudes of both of these harmonics may be set by nonlinear processes, as addressed in \se{s:anom}.  If correct, this implies that these particular modes strictly cannot be modeled using the linear methods we focus on in this paper.

We are thus justified in restricting our analysis to only those integral harmonics in the range $35\leq k \leq 89$.  We chose $k = 35$ as our lower bound to avoid modeling harmonics that contribute to ellipsoidal variation. We set $m=0$ for all of our analysis for the reason stated above. We also only used $l=2$ for the tidal potential, since additional $l$ terms are suppressed by a) further powers of $R/\Dp\sim0.16$, and b) smaller disk-integral factors from \se{s:obsflux} (e.g., $b_3/b_2=0.2$).

To find a reasonable fit to the harmonic power observed in \koi{}, we attempted a simplistic, brute-force optimization of our model against the data: we first modeled the linear response of each stellar model in our grid separately, ignoring its companion, and calculated the resulting observed flux perturbations as a function of $k$.   We then compared the absolute values of these flux perturbations to the observations of \koi{} and selected the best $10^3$ parameter choices $(M, R)$  for each star. (In future work, pulsation phases  should be modeled in addition to the amplitudes reported by \welsh, since this doubles the information content of the data; see also further discussion of phases in \se{s:qualdyn}.)  Given this restricted set of stellar models, we computed the theoretical Fourier spectra for all $10^6$ possible pairings of models.  

\begin{figure}
\centering
    $|\delta J_k/J|\times10^5$\hfill$\phantom{|\delta J_k/J|\times10^5}$\\
   \includegraphics{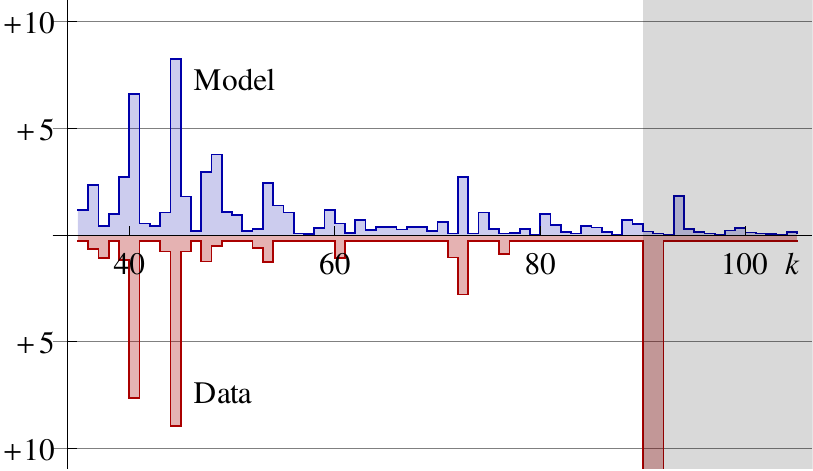}
   \caption{Exploratory modeling of the Fourier power spectrum of \koi's lightcurve. Our inhomogeneous theoretical model, including nonadiabaticity and rotation in the traditional approximation, is plotted above the graph's horizontal axis, while the data for \koi{}  (\ta{t:data}) is plotted below. Parameters corresponding to this plot are in \ta{t:bestfit}. Since harmonics 90 \& 91 represent extreme resonances, they are difficult to resolve in a given stellar model grid, and fits which reproduce their amplitudes cannot reproduce other parts of the Fourier spectrum. Thus we attempted to fit only $35\leq k \leq 89$ and not the shaded region. Our fitting process was simplistic (see text), and we did not approach a full optimization, although our best fit does agree reasonably well with the data. \fig{f:spec_fine} shows the observed flux perturbation from this plot for both stars separately on a fine frequency grid.}
   \label{f:results}
\end{figure}

\fig{f:results} shows one of our best fits to the observations of \koi{}; \ta{t:bestfit} gives the associated stellar parameters.   We obtained many reasonable fits similar to \fig{f:results} with dissimilar stellar parameters, demonstrating that many local minima exist in this optimization problem.  As a result, \fig{f:results} and \ta{t:bestfit} should not be interpreted as true best fits but rather as an example of a model that can semi-quantitatively explain the observed harmonic power in \koi{}.   We leave the task of using the observed pulsation data to quantitatively constrain the structure of the stars in \koi{} to future work, as we discuss in \se{s:conc}.

\begin{table}
\centering
\caption{Stellar parameters used in Figures \ref{f:results} and \ref{f:spec_fine}.}\label{t:bestfit}
\begin{tabular}{ccccc}\hline
 star & $M/M_\Sun$ & $R/R_\Sun$ & $Z$ & $\Prot/\mr{day}$ \\\hline
1 & 2.278 & 2.204 & 0.04 & 1.5 \\
2 & 2.329 & 2.395 & 0.04 & 1.5 \\\hline
\end{tabular}\rule{7pt}{0pt}
\end{table}

\begin{figure*}
  \begin{overpic}{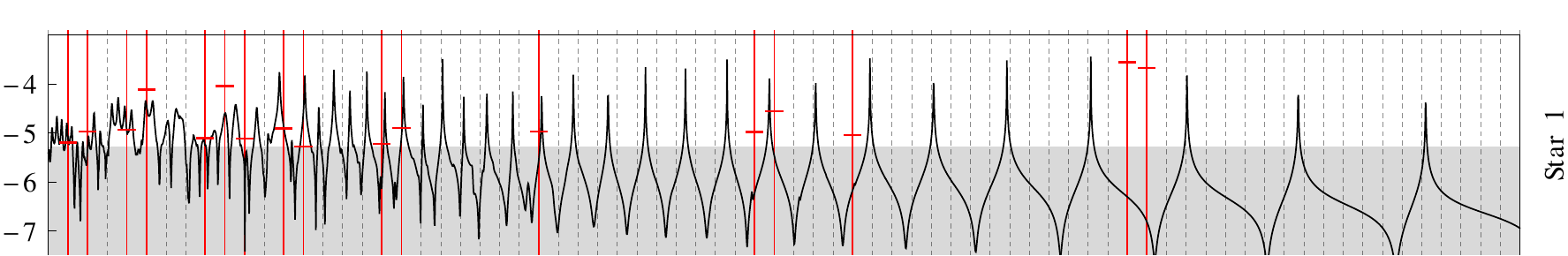}
    \put(0,15.2){$\log_{10}|\delta J_k/J|$}
  \end{overpic}
  \begin{overpic}{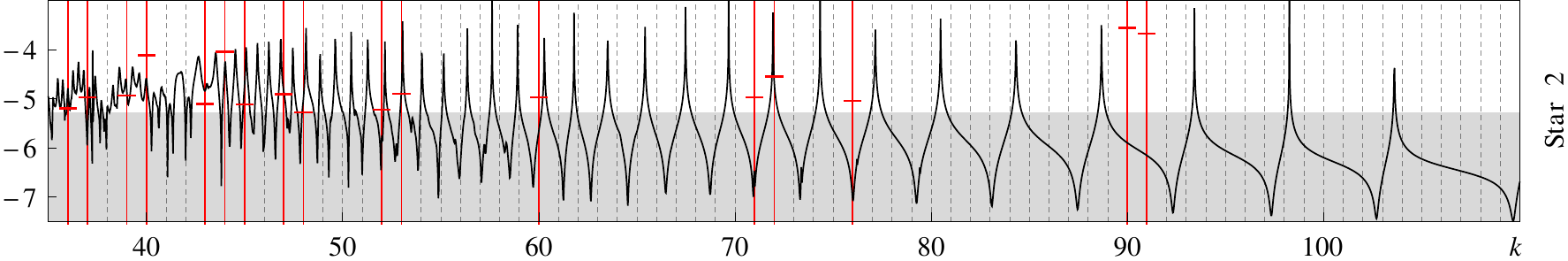}
  \end{overpic}
  \caption{Individual stars' contributions to the Fourier spectrum from \fig{f:results} (black curves), evaluated on a fine grid in $k=\omega/\omegorb$. The actual Fourier series decomposition (\fig{f:results}) is obtained by adding both stellar responses (together with phases) at each integer value of $k$. Observed harmonics are displayed as red vertical lines, with short red horizontal lines indicating their amplitudes. We used a mesh of 50 points per unity increment in $k$ to produce these plots, and find that the highest peaks of the eigenmode resonances nearest $k\sim90$ are then at the same amplitude level as the observed 90th and 91st harmonics; this means that a detuning of $\Delta k =\delta\omega/\omegorb\sim 0.02$ is required to explain these pulsations if they have $m=0$ (see text). The regions below the minimum amplitude reported by \welsh{} (\ta{t:data}) are shaded.}
  \label{f:spec_fine}
\end{figure*}

Responses from both stars were used to create the plot in \fig{f:results}.  \fig{f:spec_fine}, on the other hand, uses the same parameters, but instead shows each star's observed flux perturbation separately and evaluated on a fine grid in frequency rather than only at integral orbital harmonics. As a result, \fig{f:spec_fine} exposes the position of normal modes (which correspond to peaks in the black curves) in relation to observed harmonics (shown as red vertical lines), as well as other features not captured in \fig{f:results}'s raw spectrum. \fig{f:spec_fine} also shows that harmonics 90 and 91 must come from different stars if they are indeed $m=0$ g-modes (although this may not be the case; see \se{s:psmode}), since the g-mode spacing near $k\sim90$ is much larger than the orbital period (with the same logic applying for harmonics 71 and 72).

\subsection{Nonharmonic pulsations: three-mode coupling} \label{s:anom}
\welsh{} report nine pulsations which are not obvious harmonics of the orbital frequency; these have asterisks next to them in \ta{t:data}. As we showed previously (\se{s:pureharm} and \ap{a:zeroth}), these cannot be linearly driven modes. Here we present one possible explanation for the excitation of these pulsations.

To begin, we point out the following curious fact: the two highest-amplitude nonharmonic pulsations in \ta{t:data} (F5 and F6) have frequencies which sum to 91.00 in units of the orbital frequency---precisely the harmonic with the second-largest amplitude (F2). (This is the only such instance, as we discuss below.)

Although this occurrence could be a numerical coincidence, it is strongly suggestive of parametric decay by nonlinear three-mode coupling, the essential features of which we now  describe. First, however, we emphasize that the treatment we present here is only approximate. In reality, the process of nonlinear saturation is much more complicated, and a more complete calculation would involve fully coupling a large number of eigenmodes simultaneously \citep{weinberg08}.

If a parent eigenmode is linearly excited by the tidal potential to an amplitude that surpasses its three-mode-coupling threshold amplitude $S_a$, any energy fed into it above that value will be bled away into daughter mode pairs, each with frequencies that sum to the parent's oscillation frequency \citep{weinberg11}.   In a tidally driven system, the sum of the daughter modes' frequencies must thus be a harmonic of the orbital frequency.

For a parent with indices $a = (n,l,m)$ linearly driven at a frequency $\sigma$, the threshold is given by 
\newcommand{\detune}{\delta\omega_{bc}}
\begin{equation}\label{e:3mode}
 |S_a|^2 \simeq \mathop{\mr{min}}_{bc}\big(T_{abc}\big),
\end{equation}
where
\begin{equation}
\label{e:thresh}
 T_{abc} = \frac{\gamma_b \gamma_c}{4\omega_b\omega_c|\kappa_{abc}|^2} \left( 1+\frac{\detune^2}{\left( \gamma_a+\gamma_b \right)^2} \right),
\end{equation}
$\omega_i$ is a mode frequency, $\gamma_i$ is a mode damping rate, $\kappa_{abc}$ is the normalization-dependent nonlinear coupling coefficient \citep{schenk02}, $\detune=\sigma - \omega_b - \omega_c$ is the detuning frequency, and the minimization is over all possible daughter eigenmodes $b$ and $c$ (each short for an $(n,l,m)$ triplet).\footnote{This section uses the normalization of \citet{weinberg11}, whereas the rest of the paper uses the normalization given in \se{s:nmodes}. We of course account for this when giving observable quantities.} The nonlinear coupling coefficient $\kappa_{abc}$ is nonzero only when the selection rules
\begin{align}
 \label{e:selrules1}&0=\mr{mod}(l_a+l_b+l_c,\;2),\\
 \label{e:selrules2}&0=m_a+m_b+m_c,\\
 \label{e:selrules3}&|l_b-l_c|<\;l_a<l_b+l_c
\end{align}
are satisfied. Due to the second of these rules, any Doppler shifts due to rotation do not affect the detuning since they must cancel. 

For a simple system of three modes, the nonlinear coupling's saturation can be determined analytically.   The parent saturates at the threshold amplitude $S_a$, and the ratio of daughter energies within each pair is given by the ratio of the daughters' quality factors:
\begin{equation}
 \frac{E_b}{E_c} = \frac{q_b}{q_c} = \frac{\omega_b/\gamma_b}{\omega_c/\gamma_c}.
\end{equation} 

Equations \eqref{e:3mode} and \eqref{e:thresh} exhibit a competition that determines which daughter pair will allow for the lowest threshold. At larger daughter $l$, modes are more finely spaced in frequency, since g-mode frequencies roughly satisfy the asymptotic scaling from \eq{e:gfreq}; hence, the detuning $\detune$ becomes smaller (statistically) with increasing $l$. However, higher daughter $l$ also leads to increased damping rates at fixed frequency (\eqp{e:gamscale}). As such, the minimum threshold will occur at a balance between these two effects.

In order to semi-quantitatively address the phenomenon of three-mode coupling in \koi, we produced an example calculation of $S_a$ together with a list of best-coupled daughter pairs. To this end, we used a MESA stellar model \citep{paxton11} consistent with the mean values of star 1's properties reported in \welsh{}  (\ta{t:wparams}). We computed this model's adiabatic normal modes using the ADIPLS code \citep{dalsgaard08}, and calculated each mode's global quasiadiabatic damping rate $\gamma_{nl}$ due to radiative diffusion (\se{s:nmodes}).

We focus on the second-highest-amplitude $k=91$ harmonic present in the data (F1 from \ta{t:data}) and set $\sigma=91\times\omegorb$; as pointed out in \welsh{}, for $m_a=0$ the quadrupolar eigenmode with natural frequency closest to the 91st orbital harmonic is the g$_{14}$ mode, i.e., the g-mode with 14 radial nodes. We thus take this as our parent mode.

The minimization in \eq{e:3mode} is over all normal modes, of which there is an infinite number. To make this problem tractable numerically, we essentially followed the procedure described in \citet{weinberg08}:

\begin{enumerate}
\item We restricted daughter modes to $1\le l \le 6$. There is no reason \emph{a priori} to suggest $l$ should be in this range, but, as shown in \ta{t:daughters}, $1\leq l\leq3$ turns out to be the optimum range for minimization in this particular situation, and modes with $l>6$ are irrelevant.

\item The quantity to be minimized in \eq{e:3mode}, $T_{abc}$, achieves its minimum at fixed $\detune$ and $\kappa_{abc}$ for $\omega_b \approx \omega_c$, given the scaling from \eq{e:gamscale}. As such, we computed all normal modes $b$ with frequencies in the range $f < \omega_b/\omega_a < 1-f$; we took $f=1/10$, which yielded 344,479 potential pairs, but trying $f=1/5$, which yielded 61,623, did not change the result.

\item We computed $T_{abc}$, not including the three-mode-coupling coefficient $\kappa_{abc}$ (since it is computationally expensive to evaluate), for all possible pairs of modes satisfying (i) and (ii) as well as the selection rules in equations \eqref{e:selrules1} -- \eqref{e:selrules3}.
 
\item From the results of (iii), we selected the $N=5000$ smallest threshold energies, and then recomputed $T_{abc}$ for these pairs this time including $\kappa_{abc}$ \citep{weinberg11}. (Trying $N=1000$ did not change the results.) We set $m_b=m_c=0=m_a$ for simplicity, since $\kappa_{abc}$ depends only weakly on the values of $m$ so long as \eq{e:selrules2} is satisfied. Sorting again then yielded the best-coupled daughter pairs and an approximation for the saturation amplitude $S_a$.
\end{enumerate}

 \begin{table*}
   \caption{Ten best-coupled daughter mode pairs resulting from the procedure outlined in steps (i -- iv) of \se{s:anom}. This is an example calculation and is not meant to quantitatively predict the nonharmonic components of \koi's lightcurve. All frequencies and damping rates are in units of $\omegorb$. The square root of the daughter quality factor ratio, $\sqrt{q_b/q_c}$, gives an estimate of the ratio of daughter mode amplitudes, and hence of their potential relative lightcurve contributions.}
   \label{t:daughters}
   \begin{center}
     \begin{tabular}{lr@{ : }lr@{.}lr@{.}lr@{.}lr@{.}lr@{.}l}
       \hline
       ID&$(l_b,n_b)$&$(l_c,n_c)$ & \multicolumn{2}{c}{$\omega_b$} & \multicolumn{2}{c}{$\omega_c$} & \multicolumn{2}{c}{$\tfrac{1}{2}\log_{10}(q_b/q_c)$} & \multicolumn{2}{c}{$\log_{10}|\detune|$} & \multicolumn{2}{c}{$\log_{10}(2\sqrt{\gamma_b\gamma_c})$} \\\hline
P1 & (2, -37) & (2, -23) & 35&3 & 55&8 & \phantom{5555}0&66 & \phantom{555}-1&4 & \phantom{5555}-2&1\\
P2 & (1, -25) & (3, -30) & 29&9 & 61&0 & 0&057 & -1&5 & -2&5\\
P3 & (1, -28) & (1, -11) & 26&8 & 64&1 & 0&69 & -1&3 & -3&1\\
P4 & (1, -50) & (3, -24) & 15&3 & 75&7 & 1&2 & -1&4 & -1&7\\
P5 & (1, -27) & (3, -29) & 27&8 & 63&2 & 0&031 & -1&4 & -2&5\\
P6 & (1, -42) & (3, -25) & 18&0 & 73&1 & 1&2 & -1&6 & -1&7\\
P7 & (1, -36) & (1, -10) & 20&9 & 69&8 & 1&3 & -0&61 & -2&7\\
P8 & (2, -35) & (2, -24) & 37&2 & 53&7 & 0&51 & -0&87 & -2&2\\
P9 & (2, -29) & (2, -28) & 44&7 & 46&4 & 0&038 & -0&99 & -2&5\\
P10 & (1, -35) & (1, -10) & 21&5 & 69&8 & 1&2 & -0&51 & -2&8\\\hline
     \end{tabular}
   \end{center}
 \end{table*}

\ta{t:daughters} shows the best-coupled daughter mode pairs resulting from this procedure. It is interesting to note that most daughter pairs a) involve an $l=1$ mode coupled to an $l=3$ mode (P2, P4, P5, P6), and/or b) have a large quality-factor ratio (all except P2, P5, \& P9 have $|\tfrac{1}{2}\log_{10}(q_b/q_c)| > 0.5$).

For daughter pairs satisfying (a), the $l=3$ mode would be much harder to observe in a lightcurve since disk averaging involves strong cancellation for larger-$l$ modes---indeed, \ta{t:diskint} shows $b_3/b_1\sim 0.1$ for Eddington limb darkening, where $b_l$ is a disk-integral factor defined in \eq{e:diskb}. (The other disk-integral factor, $c_l$, does not decline as sharply with increasing $l$, but corresponds to cross-section perturbations, which are small relative to emitted flux perturbations as discussed below.)  For daughter pairs satisfying (b), since the ratio of daughter amplitudes scales as the square root of the ratio of their quality factors, one of the modes would again be difficult to observe.

Furthermore, if the parent had $m_a=\pm2$ instead of $m_a=0$ (see \se{s:psmode}), each daughter pair would have several options for $m_b$ and $m_c$, introducing the possibility of $|m_b|\ne |m_c|$. This would mean daughters would experience even greater disparity in disk-integral cancellation due to the presence of $\Yo$ in \eq{e:fluxvar}; e.g., $|Y_{10}(\theta_o,\phi_o)/Y_{32}(\theta_o,\phi_o)|\sim0.02$ for \koi.

The above results provide a reasonable explanation for why there is only one instance of two nonharmonic pulsations adding up to an observed harmonic in the data for \koi---only P9 from \ta{t:daughters} has the potential to mimic pulsations F5 and F6 from \welsh{}. Nonetheless, the nonlinear interpretation of the nonharmonic pulsations in \koi{} predicts that every nonharmonic pulsation should be paired with a lower-amplitude sister such that their two frequencies sum to an exact harmonic of the orbital frequency. This prediction may be testable given a sufficient signal-to-noise ratio, which may be possible with further observations of \koi.

Lastly, we can attempt to translate our estimate of the parent threshold amplitude $S_a$ into an observed flux perturbation, $\delta J_\mr{sat}/J$, using the techniques of \se{s:obsflux}. Since our nonlinear saturation calculation was performed with adiabatic normal modes, we strictly can only calculate the observed flux variation due to cross-section perturbations, $\delta J_\mr{cs}$ (the $\xi_r$ component of \eqp{e:fluxvar}), and not that due to emitted flux perturbations, $\delta J_\mr{ef}$ (the $\Delta F$ component of \eqp{e:fluxvar}). It evaluates to
\[\begin{split}
 \left|\frac{\delta J_\mr{cs}}{J}\right| &= \big|S_a\times(2b_l-c_l)\times\xi_{r,a}(R)\times Y_{20}(\theta_o, \phi_o)\big|\\
 &\simeq 1.7 \ \mr{mmag},
\end{split}\]
However, we can employ our nonadiabatic code to calibrate the ratio of $\delta J_\mr{cs}$ to $\delta J_\mr{ef}$, which we find to be $\delta J_\mr{ef}/\delta J_\mr{cs} \simeq 9$ for the 91st harmonic. We can then estimate the total saturated flux perturbation:
\[
 \left|\frac{\delta J_\mr{sat}}{J}\right| = \left(\frac{\delta J_\mr{ef}}{\delta J_\mr{cs}}+1\right)\left|\frac{\delta J_\mr{cs}}{J}\right| \simeq \text{17}\ \mr{mmag}.
\]

This result is a factor of $\sim$  100 too large relative to the observed amplitude of 229 $\mu\mr{mag}$ for the 91st harmonic (\ta{t:data}). Taken at face value, this would mean that the inferred mode amplitude is below threshold, and should not be subject to nonlinear processes, despite evidence to the contrary.  There are several possible explanations for this discrepancy.   If the 91st harmonic is actually an $m = \pm 2$ mode, which we proposed in \se{s:pseudo}, then the intrinsic amplitude required to produce a given observed flux perturbation is a factor of $\sim$ 200 times larger than for $m = 0$ modes given \koi{}'s face-on inclination (\se{s:obsflux}).   This would make the observed flux perturbation  of the 91st harmonic comparable to that corresponding to the threshold for three-mode coupling, consistent with the existence of nonharmonic pulsations in the lightcurve. We discuss this further in \se{s:psmode}.

Alternatively, if the 91st harmonic is in fact an $m = 0$ mode, many daughter modes may coherently contribute to the parametric resonance, reducing the threshold considerably, as in \citet{weinberg11}. A more detailed calculation, coupling many relevant daughter and potentially granddaughter pairs simultaneously, should be able to address this more quantitatively.

\subsection{Are harmonics 90 and 91 caused by prograde, resonance-locking, $\bf \boldsymbol{|m|}=2$ g-modes?}\label{s:psmode}
As introduced in \se{s:pseudo}, having two pseudosynchronized stars presents an ostensibly appealing explanation for the large-amplitude 90th and 91st harmonics observed in \koi{} (henceforth F1 and F2; \ta{t:data}): each is the manifestation of a different highly resonant eigenmode effecting a resonance lock for its respective star by opposing the equilibrium tide's torque.

We discuss the viability of this interpretation below.   First, however, what alternate explanation is available?    The most plausible would be that F1 and F2 are completely independent,  resonantly excited $m = 0$ modes.   Each coincidence would require a detuning of $|\omegnl -\dopp|/ \omegorb\sim 2\times10^{-2}$ (\fig{f:spec_fine}), which is equivalent to $|\omegnl -\dopp|/\omegnl\sim 10^{-4}$, where $\omegnl$ is the nearest eigenfrequency and $\dopp=k\omegorb-m\omegrot$ is the driving frequency. The probability of having a detuning equal to or smaller than this value, given $\sim$ 10 available modes (\fig{f:spec_fine}), is $\sim$ 10\%, so the combined probability if the resonances are independent is $\sim$ 1\%.  Moreover, in \se{s:model} we show that in this $m=0$ interpretation, F1 and F2 must come from different stars, yet there is no explanation for why the two excited modes are so similar.

If instead F1 and F2 are due to highly resonant $m = \pm 2$ resonance locking modes, several observations are naturally explained.  The high degree of resonance is an essential feature of the inevitable pseudosynchronous state reached when the torque due to the dynamical tide cancels that due to the equilibrium tide (\se{s:pseudo}).   The fact that the resonant modes correspond to similar $k$ would be largely a consequence of the fact that the two stars in the \koi{} system are similar in mass and radius to $\sim$ 10\%, so that a similar mode produces the dynamical tide torque in each star (although a corresponding $\sim$ 10\% difference in $k$ would be equally possible in this interpretation).

In addition, we showed in \se{s:anom} that the observed amplitudes of F1 and F2  are  a factor of up to $\sim$ 100 smaller than their nonlinear threshold values assuming $m=0$. There is also strong evidence that at least F2 has its amplitude set by nonlinear saturation.  Having $m\ne0$ would help to resolve this discrepancy because the intrinsic amplitude of $m = \pm 2$ modes would need to be $\sim$ 200 times larger to produce the observed flux perturbation.   This would then imply that the amplitudes of F1 and F2 are indeed above the threshold for three-mode coupling, naturally explaining the presence of the nonharmonic pulsations in the \koi{} lightcurve.

However, several significant problems with the resonance-locking interpretation arise upon closer examination.  Assume that F1 and F2 indeed correspond to $m=\pm2$ g-modes that generate large torques effecting $\Ppsp\sim 1.8$ days pseudo\-syn\-chron\-ization locks. In order to create positive torques, \eq{e:avetau} shows that we must have $m(k\omegorb-m\omegrot)>0$, which reduces to
 \begin{equation}\label{e:prog}
  (k/m)\omegorb>\omegrot.
 \end{equation}
In order to determine which modes correspond to F1 and F2, we can enforce a close resonance by setting
 \begin{equation}\label{e:progres}
  \omegnl\simeq 90\,\omegorb-2\,\omegrot,
 \end{equation}
where we have used the fact that \eq{e:prog} requires $k$ and $m$ to have the same sign for a positive torque. For $l=2$ and using a MESA stellar model consistent with \welsh's mean modeled parameters for star 1 (\ta{t:wparams}), \eq{e:progres} yields $n\simeq30$, neglecting rotational modification of the modes (i.e., not employing the traditional approximation).
 
However, in our calculations in \se{s:pseudo} we find that the resonant torque due to the dynamical tide is instead typically caused by g-modes with $n$ of 8 -- 15 (basically set by the intersection of the Hansen coefficient and linear overlap curves, as discussed in \se{s:qualdyn} in the context of flux perturbations). Using \eq{e:progres} again, this would mean we would expect $k$ of 140 -- 200.   Furthermore, we find that even a perfectly resonant $n=30$ g-mode makes a negligible contribution to the torque.   This is true both for ZAMS models and for evolved models consistent with the observed radii in \koi{}, indicating that there is little uncertainty introduced by the details of the stellar model.   This result suggests that the g-modes inferred to correspond to F1 and F2 are inconsistent with what would be expected from our torque calculation if the rotation rate is indeed $\sim$ 1.8 days.
 
If we account for rotation in the traditional approximation (\se{s:rot}), the $n$ of a prograde mode of a given frequency can be at most a factor of $\sqrt{4/6}$ times smaller than its corresponding nonrotating value; this follows from the fact that the angular eigenvalue $\lambda$ asymptotes to $m^2 = 4$ in the limit $\omegnl \ll \Omega_*$ for prograde modes, instead of $\lambda = l(l + 1)=6$ in the nonrotating limit. This reduces the $n$ of the $90$th harmonic from $n \simeq 30$ to $n \simeq 24$, still insufficient to yield a significant torque.

Another major problem with the resonance lock interpretation is that although our orbital evolution simulations described in \se{s:pseudo} ubiquitously produce resonance locks, they always occur in only one star at a time. This is because if a mode is in a resonance lock in one star and a mode in the other star tries to simultaneously resonance lock, the first lock typically breaks since the orbital frequency begins to evolve too quickly for the lock to persist. Although it is possible for simultaneous resonance locks in both stars to occur, such a state is very improbable. Similar orbital evolution simulations presented in \citet{fuller11} did produce simultaneous resonance locks, but only because they simulated only one star and simply doubled the energy deposition rate and torque, thus not allowing for the effect just described.
 
Finally, we point out one last inconsistency in the resonance lock interpretation of F1 and F2. It is straightforward to calculate the predicted flux perturbation associated with perfectly resonant $|m| = 2$ g-modes in resonance locks (using, e.g., the calibration discussed at the end of \se{s:anom}): for modes ranging from $n \sim$~8 -- 15, we find that the predicted flux perturbation for \koi{}'s parameters is $\sim$ 10 -- 30 $\mu\text{mag}$.   This is a factor of $\sim$ 10 smaller than the observed flux perturbations, yet somewhat larger than the smallest-amplitude pulsation reported by \welsh. It is also a factor of $\sim 2$ smaller than the nonlinear coupling threshold for an $m=2$ mode (which we determined using the same procedure as in \se{s:anom}, extended to allow for an $m\ne0$ parent), although the uncertainties involved in our nonlinear estimates are significant enough that we do not consider this to be a substantial problem.

Thus even if F1 and F2 can be attributed to modes undergoing resonance locks (which is highly nontrivial, as we have seen), the observed amplitudes are larger than those we predict. Conversely, if F1 and F2 are simply chance $m=0$ resonances, it appears that if a resonance lock existed, it would have been detected, although the possibility exists that the resonant mode's flux perturbation was marginally smaller than those of the 30 reported pulsations due to uncertainties in our calculations. Firmer constraints on the flux perturbations in \koi{} at $k\sim$ 140 -- 200 would be very valuable in constraining the existence of such $m = \pm 2$ modes, as would information about the phases of the 90th and 91st harmonics (see \se{s:qualdyn}).

\section{Discussion}\label{s:conc}
We have developed a set of theoretical tools for understanding and modeling photometric observations of eccentric stellar binaries. This work is motivated by the phenomenal photometry of the Kepler satellite and, in particular, by the discovery of the remarkable eccentric binary system \koi{} (\citealt{welsh11}; henceforth \welsh). This system consists of two similar A stars exhibiting strong ellipsoidal lightcurve variation near  periastron passage due to the system's large ($e=0.83$) eccentricity. \welsh{} successfully modeled this phenomenon, and also reported the detection of at least 30 distinct sinusoidal pulsations in \koi's lightcurve (\se{s:koibg}), $\sim$ 20 at exact harmonics of the orbital frequency and another $\sim$ 10 nonharmonic pulsations. Although our work has focused on modeling \koi, our methods and techniques are more general, and are applicable to other similar systems.

We developed a simple model of \koi's periastron brightening, including both the irradiation and equilibrium tide components of this effect, which agrees at the $\sim$ 20\% level with the results \welsh{} obtained using a much more detailed simulation (\se{s:ellips}). Our model may be useful for analysis of other eccentric stellar binaries, allowing determination of orbital and stellar parameters; its simplicity should enable it to be implemented in an automated search of Kepler data.

In \se{s:qualres} we used the adiabatic normal mode formalism (see \se{s:nmodes} and, e.g., \citealt{dalsgaardnotes,kumar95}), to establish a qualitative connection between the range of stellar modes excited in a given binary system and the system's orbital properties. For more detailed quantitative modeling of the harmonic pulsation spectrum of a given binary system, we further developed the nonadiabatic, inhomogeneous tidal method from \citet{pfahl08} by including the Coriolis force in the traditional approximation (\se{s:nonad}; \ap{a:nonad}).

In \se{s:rotres} we used this method to show that  fast rotation tends to suppress power in the lower harmonics of a face-on binary system's lightcurve (\fig{f:rotres}).   This can qualitatively explain why there is a scarcity of large-amplitude, lower-harmonic pulsations in \koi{}'s lightcurve, relative to predictions for nonrotating stars (\fig{f:hansoverlapprod}). We also showed in \se{s:rotres}, however, that the dynamical tidal response may be much larger than ellipsoidal variation in edge-on binaries, unlike in \koi{} (which has an inclination of $i=5.5\deg$; see \ta{t:wparams}). For such systems, simultaneous modeling of the dynamical and equilibrium tides may be required in order to constrain system properties.

Moreover, in \se{s:sync} we showed that rapid rotation periods of $\sim$ 1.8 days are expected for the A stars in \koi{}, due to pseudosynchronization with the orbital motion near periastron.   This pseudosynchronous rotation period is shorter than the value of $2.53$ days assumed by \welsh.   The latter value is appropriate if the only appreciable torque is that produced by the equilibrium tide (\ap{a:pseudo}). Since resonantly excited stellar g-modes can produce a torque comparable to that of the equilibrium tide, pseudosynchronous rotation can occur at even shorter rotation periods (\fig{f:torque}). This involves a stochastic equilibrium between prograde resonance locks and the equilibrium tide. These same rapid rotation periods ($\sim$ 1.8 days) yield predicted lightcurve power spectra that are the most qualitatively consistent with the pulsation data for \koi{} (\fig{f:rotres}).
 
In \se{s:model} we performed a preliminary optimization of our nonadiabatic model by comparing its results in detail to the Fourier decomposition of \koi's lightcurve (\ta{t:data}).  We searched over an extensive grid of stellar masses and radii, assuming a metallicity of twice solar and a rotation period of $1.5$ days.  We also set $m=0$, since \koi's nearly face-on orientation implies that this is the case for almost all of the pulsations we modeled (\se{s:obsflux}). The modeling challenge in tidal asteroseismology contrasts with that of standard asteroseismology in that a) we must simultaneously model both stars, and b) pulsation amplitudes and phases  contain the key information in our case, since we are considering a forced system, whereas pulsation frequencies constitute the data in traditional asteroseismology.   Moreover, stellar rotation is sufficiently rapid in eccentric binaries that its effect on stellar g-modes cannot be treated perturbatively.

Although our minimization procedure was quite simple, we were able to obtain stellar models with power spectra semi-quantitatively consistent with the observations of \koi{} (\fig{f:results} \& \fig{f:spec_fine}).  The resulting model in \fig{f:results} is not formally a good fit, but this is not surprising given that two of the key parameters (metallicity and rotation period) were not varied in our analysis.   Moreover, in our preliminary optimization we found that there were many local minima that produced comparably good lightcurves.

As noted above, {\em a priori}  calculations suggest that both stars in the \koi{} system should have achieved a pseudosynchronous state at rotation periods of $\sim$ 1.8 days. This requires frequent resonance locks to occur, when a single $|m| = 2$ eigenmode comes into a near-perfect prograde resonance.   A natural question is whether such a highly resonant mode could contribute to the \koi{} lightcurve; this possibility is particularly attractive for the two largest-amplitude harmonics observed, the 90th and 91st. (See also our calculation of nonlinear saturation from \se{s:anom}, discussed below.)

However, we find quantitative problems with this interpretation (\se{s:psmode}).   First, our orbital evolution simulations (\se{s:pseudo}) indicate that only one resonance lock should exist at a time, meaning that only one of the two large-amplitude harmonics could be explained in this way. This result is in disagreement with the simulations performed by \citet{fuller11}, since they did not simultaneously model both stars.

Further, in our calculations, the g-modes capable of producing torques large enough to effect resonance locks have $n$ typically in the range of 8 -- 15 (where $n$ is the number of radial nodes), while the $90$th harmonic corresponds to $n$ of 25 -- 40 for $m =\pm2$ and rotation periods of 2.0 -- 1.5 days.   Also, we predict that g-modes producing resonance locks should have  $k$ of 140 -- 200, much larger than $\sim$ 90, and flux perturbations of 10 -- 30 $\mu{\rm mag}$. The latter values are a factor of $\sim$ 10 less than that observed for the 90th and 91st harmonics, but slightly larger than the smallest observed pulsations.

It thus seems quantitatively difficult to interpret harmonics 90 and 91 in \koi{} as manifestations of $m = \pm2$ modes in resonance locks, although we cannot conclusively rule out this possibility. Instead, it seems likely that they are simply chance $m=0$ resonances (as is almost certainly the case for the overwhelming majority of the other observed pulsations in \koi). One theoretical uncertainty resides in our omission of rotational modification of the stellar eigenmodes when computing tidal torques. Our estimates suggest that this is a modest effect and is unlikely to qualitatively change our conclusions, but more detailed calculations are clearly warranted.

We note that in future work, pulsation phases  should be modeled in addition to the amplitudes reported by \welsh, since this effectively doubles the information content of the data. Indeed, we showed in \se{s:qualdyn} that a resonant pulsation's phase is strongly influenced by the mode's value of $m$. In particular, since harmonics 90 and 91 are likely standing waves, as can be seen in the propagation diagram in \fig{f:prop}, measurement of their phases could help to resolve the uncertainties pointed out above by supplying direct information about their degrees of resonance, thus potentially confirming or disproving the $m=\pm2$ resonance lock  interpretation.
 
In \se{s:anom} we pointed out evidence for nonlinear mode coupling in \koi's observed pulsations: the existence of nonharmonic pulsations (which does not accord with linear theory; \se{s:pureharm}) and the fact that two of them have frequencies that sum to exactly the frequency of the 91st harmonic, the second-largest-amplitude harmonic pulsation in \koi's lightcurve.   This is consistent with parametric resonance, the leading-order nonlinear correction to linear stellar oscillation theory \citep{weinberg11}.

Motivated by this observation, we performed a nonlinear stability calculation that qualitatively explains why no other similar instance of a nonharmonic pair summing to an observed harmonic is present in the data:   for the majority of daughter pairs likely to be nonlinearly excited, there are sufficient differences in the $l$ and $m$ values of the daughter pair members, or sufficient differences in their predicted saturated energies, that only one member of the pair would be observable given current sensitivity.   Nonetheless, the nonlinear interpretation makes the strong prediction that  every nonharmonic pulsation should be paired with a lower-amplitude sister such that their two frequencies sum to an exact harmonic. This prediction may well be testable given a better signal-to-noise ratio.

One additional feature of the nonlinear interpretation is that if the nonlinearly unstable parent is an $m = 0$ mode, then the threshold amplitude for a linearly excited mode to be unstable to parametric resonance, which we have just argued exists in \koi, implies flux perturbations that are a factor of $\sim$ 100 larger than those observed.   In contrast, the parent being an $m = \pm2$ mode ameliorates this discrepancy because the parent's intrinsic amplitude must be $\sim$ 200 times larger for a given flux perturbation due to \koi's face-on orientation.   This result thus argues in favor of the 91st harmonic in \koi{} being an $m = \pm2$ mode caught in a resonance lock, as discussed above.

There are many prospects for further development of the analysis begun in this paper. For example, in traditional asteroseismology, standard methods have been developed allowing a set of observed frequencies to be inverted uniquely, yielding direct constraints on stellar parameters, including the internal sound speed profile \citep{unno89}. The essential modeling difficulty in tidal asteroseismology is our inability to assign each observed pulsation amplitude to either star of a given binary \emph{a priori}, hindering our attempts to develop a direct inversion technique. We leave the existence of such a technique as an open question.

Future observations of eccentric binaries may avoid this difficulty if one star is substantially more luminous than the other. However, for eccentric binaries with similar stars, in the absence of a means of direct inversion, we are left with a large parameter space over which to optimize, consisting at minimum of eight quantities: both stars' masses, radii/ages, metallicities, and rotation periods. Even this parameter set may ultimately prove insufficient, if modeling of tidally forced pulsations is found to be sensitive to the details of e.g.\ chemical mixing or convective overshoot, which can modify the \brunt{} frequency and thus g-mode frequencies.

One possible approach that should be explored in future work is to apply standard numerical optimization algorithms such as simulated annealing to this parameter space, attempting to minimize the $\chi^2$ of our nonadiabatic code's theoretical Fourier spectrum against the observed harmonic pulsation data.  In practice, it may be preferable to develop interpolation techniques over a grid of models given the high resolution in stellar parameters needed to resolve the close resonances responsible for large-amplitude pulsations.

Although \koi's stars lie near the instability strip, this fact is unimportant for the tidal asteroseismology theory presented in this work. Consequently, future high-precision photometric observations of other eccentric binaries may supply a window into the structure of stars previously inaccessible by the techniques of asteroseismology. Constructing a data pipeline capable of reliably flagging eccentric binary candidates---e.g., finding efficient ways of searching for the equilibrium tide/irradiation lightcurve morphologies shown in \fig{f:ellips} (\ap{a:ellips})---is also an important, complementary prospect for future work.

\acknowledgments
We are pleased to thank Geoff Marcy for several useful discussions. This work was supported by NSF AST-0908873 and NASA NNX09AF98G. J.B. is an NSF Graduate Research Fellow. E.Q. was supported in part by the David and Lucile Packard Foundation. P.A. is an Alfred P. Sloan Fellow, and received support from the Fund for Excellence in Science and Technology from the University of Virginia.   
\vspace{.4cm}

\appendix
\onecolumn

\section{Nonadiabatic Tidally Driven Oscillation Equations}\label{a:nonad}
Here we will describe the computational procedure we employed to solve for tidally driven stellar responses, which we then used to model \koi's lightcurve. In \ap{a:zeroth}, we account for rotation only by using Doppler-shifted driving frequencies $k \omegorb - m \omegrot$, and neglect any effects of the Coriolis force; in \ap{a:trad}, we invoke the traditional approximation \citep{bildsten96} to account for the Coriolis force (\se{s:rot}).

\subsection{Formalism without the Coriolis force}\label{a:zeroth}
The gravitational potential due to the secondary, experienced by the primary, is given by
\begin{equation}
U_{2\rightarrow1} = -\frac{G M_2}{|\vec{D}-\vec{r}|}.
\end{equation}
Performing a multipole expansion \citep{jackson} and excising the $l=0$ (since it is constant) and $l=1$ (since it is responsible for the Keplerian center-of-mass motion) terms, we are left with the tidal potential:
\begin{equation}\label{e:Uexpnohans}
U = -\frac{G M_2}{D(t)}\qsphsum W_{lm}\left(\frac{r}{D(t)}\right)^l e^{-imf(t)} \Ylm(\theta, \phi),
\end{equation}
where
\begin{equation}\begin{split}\label{e:W}
W_{lm} &= \frac{4\pi}{2l+1} \Ylm^*(\pi/2,0) \\
&= (-1)^{(l+m)/2}\mr{mod}(l+m+1,2)\sqrt{\frac{4\pi}{2l+1} \frac{(l+m-1)!!}{(l+m)!!} \frac{(l-m-1)!!}{(l-m)!!}}.
\end{split}\end{equation}
Next, we shift to the primary's corotating frame (by sending $\phi\rightarrow\phi+\omegrot t$) and expand the time dependence of the orbit in terms of the Hansen coefficients:
\begin{equation}\label{e:Uexp}
U= \frac{M_2}{M_1}\sum_l\left(\frac{R_1}{a}\right)^{l+1}\sum_m W_{lm}\Ylm(\theta, \phi)\sum_k \exp(-i\dopp t) \hanse U_l(r),
\end{equation}
with $\dopp = k\omegorb - m\omegrot$ and
\begin{equation}
U_l(r) = -\left(\frac{GM_1}{R_1}\right)\left(\frac{r}{R_1}\right)^l.
\end{equation}
The unit-normalized Hansen coefficients $\hansnorm$ were defined in \eq{e:hans}; here we are using the conventionally normalized Hansen coefficients $\hans = \hansnorm/(1-e)^{l+1}$, which are convenient to evaluate numerically as an integral over the eccentric anomaly:
\begin{equation}\label{e:hansexp}
\hans = \frac{1}{\pi}\int_0^\pi(1-e\cos E)^{-l}\cos\left[k(E-e\sin E) - 2 m \arctan\left(\sqrt{\frac{1+e}{1-e}}\tan(E/2)\right)\right]dE.
\end{equation}

If we represent the linear response of a star to the perturbing tidal potential by an abstract vector $\vec{y}(r,\theta, \phi, t)$ whose components are the various oscillation variables (e.g., $\xi_r/r$), then $\vec{y}$ can also be expanded, again \emph{in the primary's corotating frame,} as in \eqref{e:Uexp}:
\begin{equation}\label{e:yexp}
\vec{y}= \frac{M_2}{M_1}\sum_l \left(\frac{R_1}{a}\right)^{l+1}\sum_m W_{lm}\Ylm(\theta, \phi)\sum_k \exp(-i\dopp t) \hanse\vec{y}_{l m}^k(r).
\end{equation}
The equations necessary to determine $\vec{y}_{l m}^k(r)$ are given in the appendix of \citet{pfahl08}, along with appropriate boundary conditions; note that their $U$ is our $U_l$ and their driving frequency $\omega$ is our $\dopp$.

After determining $\vec{y}_{l m}^k(r)$ in the corotating frame, we can switch to the inertial frame specified in \se{s:prelim}:
\begin{equation}\label{e:yexpinert}
\vec{y}=\frac{M_2}{M_1}\sum_l \left(\frac{R_1}{a}\right)^{l+1}\sum_k \exp(-ik\omegorb t) \sum_m W_{lm}\Ylm(\theta, \phi)\hanse\vec{y}_{l m}^k(r).
\end{equation}
As noted in \se{s:pureharm}, we see in \eq{e:yexpinert} that the observed \emph{frequencies} should be pure harmonics of the orbital frequency, even though the corresponding \emph{amplitudes} of observed pulsations are influenced by the star's rotation rate (via the Doppler-shifted frequency $\dopp$).

\subsection{Rotation in the traditional approximation}\label{a:trad}
\newcommand{\Ullam}{U_{\lambda l m}^k}
\newcommand{\elmla}{e^q_{\lambda lm}}
\newcommand{\elmlak}{e^k_{\lambda lm}}
\newcommand{\Hlamk}{H_{\lambda m}^k}
\newcommand{\U}{U}

We now invoke the traditional approximation (\se{s:rot}); we must correspondingly adopt the Cowling approximation and employ the Hough functions (\se{f:ellips}) as angular basis functions instead of spherical harmonics.

We expand the Hough functions as \citep{longuet68}
\begin{equation}
\Hlamk = \sum_l \elmlak \tilde{P}_{lm}\quad\rightarrow\quad \elmlak  = 2\pi\int_{-1}^1 \widetilde{P}_{lm} \Hlamk d\mu\quad\rightarrow\quad\widetilde{P}_{lm} = \sum_\lambda \elmlak \Hlamk,
\end{equation}
where $\widetilde{P}_{lm}$ is a normalized associated Legendre function defined by
\begin{equation}
\widetilde{P}_{lm} = \sqrt{\frac{2l+1}{4\pi}\frac{(l-m)!}{(l+m)!}}\;P_{lm}\quad\rightarrow\quad Y_{lm}(\theta,\phi)=e^{i m \phi} \widetilde{P}_{lm}(\cos \theta).
\end{equation}
We used the numerical method of calculating the expansion coefficients $\elmlak$ detailed in \citet{ogilvie04} \S~5.4.  The tidal potential in the corotating frame is then 
\begin{equation}
U= \frac{M_2}{M_1}\sum_l\left(\frac{R_1}{a}\right)^{l+1}\sum_m W_{lm}e^{im\phi}\sum_k \exp(-i\dopp t) \hans\sum_\lambda \Hlamk(\mu)\Ullam(r),
\end{equation}
where 
\begin{equation}
\Ullam(r) = -\left(\frac{GM_1}{R_1}\right)\left(\frac{r}{R_1}\right)^l\elmlak,
\end{equation}
$\dopp = k\omegorb - m\omegrot$, and the Coriolis parameter $q$ on which the Hough functions depend is
\begin{equation}
q=2\omegrot/\dopp,
\end{equation}
which justifies writing $\Hlamk$ and $\elmlak$ rather than $\Hlam$ and $\elmla$.

We again represent the linear response of a star, as in \ap{a:zeroth}, by a vector $\vec{y}(r,\theta, \phi, t)$ whose components are the various oscillation variables, and which can be expressed in the inertial frame as:
\begin{equation}\begin{split}\label{e:yexpinerttrad}
\vec{y}&=\frac{M_2}{M_1}\sum_k \exp(-ik\omegorb t) \sum_l \left(\frac{R_1}{a}\right)^{l+1}\sum_m W_{lm}e^{im\phi}\hanse\sum_\lambda\Hlamk(\mu)\vec{y}_{\lambda l m}^k(r)\\
&=\frac{M_2}{M_1}\sum_k \exp(-ik\omegorb t) \sum_l \left(\frac{R_1}{a}\right)^{l+1}\sum_m W_{lm}\hanse\sum_\lambda\sum_{l'}e_{\lambda l'\!m}^k \,Y_{l'\!m}(\theta,\phi)\,\vec{y}_{\lambda l m}^k(r).
\end{split}\end{equation}
The expansion of $\Hlam$ back into associated Legendre functions in the second line of \eq{e:yexpinerttrad} is useful since disk integrals are convenient to perform over spherical harmonics (\se{s:obsflux}).

Following \citet{unno89}, we choose the components of $\vec{y}$ as
\begin{equation}
y_1=\frac{\xi_r}{r},\qquad y_2 = \frac{\delta p}{\rho g r}, \qquad y_5 = \frac{\Delta s}{c_p}, \quad \mr{and}\quad y_6 = \frac{\Delta L}{L_r},
\end{equation}
where we have omitted the variables corresponding to the perturbed gravitational potential, $y_3$ and $y_4$. Equation~\eqref{e:yexpinerttrad} together with determination of the radial displacement $\xi_r/r=y_1$ and the Lagrangian flux perturbation $\Delta F/F=y_6-2y_1$ at the photosphere then enables use of the formalism from \se{s:obsflux} to compute the flux perturbation as seen by an observer.

Next, we present the differential equations which determine a particular component $\vec{y}_{\lambda l m}^k(r)$ of the full response in radiative zones. These equations are nearly identical to those in the appendix of \citet{pfahl08}, but with $l(l+1)$ replaced by $\lambda$ and with certain terms set to zero as per the traditional approximation. In practice these terms can be left in, since they are nearly zero for situations where the traditional approximation is valid; this is then a smooth way of transitioning among different regimes. Omitting $(\lambda lm k)$ indices and denoting $U=\Ullam$ and $\omega=\dopp$, the equations are
\begin{align}\label{e:y1eq}
\frac{dy_1}{d\ln r} &= y_1\left(\frac{gr}{c_s^2}-3\right) + y_2\left(\frac{\lambda g}{\omega^2 r} - \frac{gr}{c_s^2}\right) - y_5 \rho_s + \frac{\lambda}{\omega^2 r^2}\U\\
\frac{dy_2}{d\ln r} &= y_1 \left(\frac{\omega^2-N^2}{g/r}\right) + y_2 \left(1-\eta+\frac{N^2}{g/r}\right) - y_5 \rho_s - \frac{1}{g} \frac{d\U}{dr}\\
\begin{split}
\frac{dy_5}{d\ln r} &= 
	y_1 \frac{r}{H_p}\left[\delad\left(\eta - \frac{\omega^2}{g/r}\right) + 4(\del - \delad)+c_2 \right]
	+ y_2 \frac{r}{H_p}\left[(\delad - \del)\frac{\lambda g}{\omega^2 r}-c_2\right]\\
	&+ y_5\frac{r}{H_p}\del (4-\kappa_s)
	- y_6\frac{r}{H_p}\del
	+\frac{r}{H_p}\left[\delad\left(\frac{d\U/dr}{g}\right)+(\delad-\del)\frac{\lambda}{\omega^2r^2}U\right]
\end{split}\\
\label{e:y6eq}\frac{dy_6}{d\ln r} &= y_2\left(\frac{\lambda \gamma g}{\omega^2 r}\right) + y_5\left(i\omega\frac{4\pi r^3 \rho c_p T}{L}\right) -y_6\gamma+\left(\frac{\lambda \gamma}{\omega^2 r^2}\right)\U
,
\end{align}
where $\eta=4\pi r^3 \rho /M_r$, $\gamma = 4\pi r^3\rho\epsilon/L_r$, $c_2 = (r/H_p)\del(\kappa_{\mr{ad}}-4\delad)+\delad(d\ln \delad/d\ln r+r/H_p)$, $H_p=\rho g/p$ is the pressure scale height, and $\epsilon$ is the specific energy generation rate.

We need four boundary conditions for our four variables. Our first three are
\begin{align}
 &&&& 0 &= \xi_r(0)	&& \text{evanescence in convective core,}&&&&\\
 &&&& 0 &= \Delta s(0)	&& \text{adiabaticity/evanescence in core,}&&&&\\
 &&&& 0 &= \frac{\Delta F(R)}{F(R)} - 4\;\frac{\Delta T(R)}{T(R)} & &\text{blackbody at the stellar surface,}&&&&
\end{align}
where $\Delta T/T$ can be cast in terms of $y_1$, $y_2$, and $y_5$ using standard thermodynamic derivative identities.

A final surface boundary condition that allows for traveling and/or standing waves can be generated by imposing energetic constraints at the surface. This is detailed in \citet{unno89} pp.~163 -- 167 for adiabatic oscillations. To generalize the boundary condition to include nonadiabaticity, rotation, and inhomogeneous tidal forcing, we write equations \eqref{e:y1eq} -- \eqref{e:y6eq} as
\begin{equation}
\frac{d\vec{y}}{d\ln r} = M\vec{y} + \vec{b},
\end{equation}
where $M$ and $\vec{b}$ are treated as constant near the stellar photosphere. The constant solution is $\vec{y}_0=-M^{-1}\vec{b}$; defining $\vec{z}=\vec{y}-\vec{y}_0$, the homogeneous solutions for $\vec{z}$ can be computed by diagonalizing $M$. In the evanescent case, we eliminate the solution for $\vec{y}$ with outwardly increasing energy density. Alternatively, in the traveling wave case, we eliminate the inward-propagating wave. The final boundary condition is then implemented by setting the amplitude of the eliminated homogeneous solution to zero, and solving for a relationship between the original fluid variables implied by this statement.

\section{Analytic Model of Ellipsoidal Variation}\label{a:ellips}
As discussed in \se{s:ellips}, our simplified model of ellipsoidal variation reproduces the much more sophisticated simulation code employed by \citet{welsh11} to model \koi; here we discuss the details of our analytic methods, which can easily be applied to model other systems.

\subsection{Irradiation}\label{a:reflect}
The following is our simple analytical model of the insolation component of the \koi's ellipsoidal variation.   We focus our analysis on the primary, since extending our results to the secondary is trivial. Our main assumption is that all radiation from the secondary incident upon the primary is immediately reprocessed at the primary's photosphere and emitted isotropically (i.e., absorption, thermalization, and reemission). This assumption is well justified for \koi, since its two component stars are of very similar spectral type.    The method below might need to be modified if the components of a binary system had significantly different spectral types, because then some of the incident radiation might instead be scattered.

\newcommand{\Fin}{F_{2\rightarrow1}}

The incident flux on the primary, using the conventions and definitions introduced in \se{s:prelim}, is
\begin{equation}
  \Fin=\frac{L_2}{4\pi D^2}\,Z(\uvec{r}\cdot\uvec{D}),
\end{equation}
where $Z$ is the ramp function, defined by
\begin{equation}
    Z(x)=\ \begin{cases}\ 0&x<0\\
    \ x & x\geq 0
  \end{cases}\ .
\end{equation}
We can expand $Z(\uvec{r}\cdot\uvec{D})$ in spherical harmonics as
\begin{equation}
  Z(\uvec{r}\cdot\uvec{D}) = \sum_{lm}Z_{lm} Y_{lm}(\theta,\phi)\,e^{-imf(t)},
\end{equation}
where $Z_{lm}$ can be evaluated to
\begin{equation}\label{e:Z}
  Z_{lm}= 2\left(\frac{2l+1}{4\pi} \cdot \frac{(l-m)!}{(l+m)!}\right)^{1/2}\left(\frac{\cos(m\pi/2)}{1-m^2}\right) \int_{-1}^1 \sqrt{1-\mu^2}P_{lm}\, d\mu,
\end{equation}
with $\cos(m\pi/2)/(1-m^2)\rightarrow \pi/2$ for $m=\pm1$.

Next, taking the reemission as isotropic, the reemitted intensity will be
\begin{equation}
  I_{\mathrm{emit}} = \frac{\Fin}{\pi}.
\end{equation}
Using this together with our expansion of $Z(\uvec{r}\cdot\uvec{D})$ as well as results from \se{s:obsflux}, we can evaluate the observed flux perturbation:
\begin{equation}\label{e:refl}
\frac{\delta J}{J_1} = \beta(T_1)\left(\frac{L_2}{L_1}\cdot\frac{R_1^2}{D(t)^2}\right)\sum_{l=0}^\infty \sum_{m=-l}^l b_{l}\, Z_{lm} \,\Yo\,e^{-imf(t)},
\end{equation}
where $ J_1 = L_1/4\pi s^2 $ is the unperturbed observed flux, $s$ is the distance to the observer, the bandpass correction coefficient $\bandpassT$ is defined in \eq{e:bandpass}, the disk-integral factor $b_l$ is defined in \eq{e:diskb}, several values of $b_l$ using Eddington limb darkening are given in \ta{t:diskint}, and other variables are defined in \se{s:prelim}. Since $b_l$ declines rapidly with increasing $l$, it is acceptable to include only the first few terms of the sum in \eq{e:refl}. We have neglected limb darkening, so it is formally necessary to use a flat limb darkening law in calculating $b_l$ ($h(\mu)=2$; \se{s:obsflux}). However, we found this to be a very modest effect.

The binary separation $D$ and the true anomaly $f$ can be obtained as functions of time in various ways, e.g., by expanding with the Hansen coefficients employed earlier (\eqp{e:hans} or \ref{e:hansexp}), or by using
\be
D = \frac{a(1-e^2)}{1+e\cos f}
\ee
together with numerical inversion of
\be
\omegorb t = 2\arctan\lp\frac{(1-e)\tan(f/2)}{\sqrt{1-e^2}}\rp - \frac{e\sqrt{1-e^2} \sin f}{1+e\cos f},\quad -\pi<f<\pi.
\ee
The observed flux perturbation from the secondary is obtained from \eq{e:refl} by switching $1\leftrightarrow2$ and sending $\phi_o\rightarrow \phi_o+\pi$.

Using the fact that $b_0=1$ and $Z_{00} = \sqrt{\pi}/2$, it can be readily verified that the total reflected power, i.e. $s^2J_1$ times \eqref{e:refl} integrated over all observer angles $ (\theta_o, \phi_o) $, is equal to $ L_2(\pi R_1^2/4\pi D^2) $. This is just the secondary's luminosity times the fraction of the secondary's full solid angle occupied by the primary, which is the total amount of the secondary's radiation incident on the primary.

\subsection{Equilibrium tide}\label{a:eqtide}
We invoke the Cowling approximation (well satisfied for surface values of perturbation variables), and use the analytic equilibrium tide solution, where the radial displacement at star 1's surface becomes
\begin{equation}
 \xi_r = -\frac{U(R_1,t)}{g(R_1)},
\end{equation} 
and $U$ is the tidal potential. Using the expansion in \eq{e:Uexpnohans},
$\xi_{r,lm}(t)/R_1$ from \eq{e:xiexpflux} becomes \citep{goldreich89}
\begin{equation}
 \frac{\xi_{r,lm}(t)}{R_1} = \frac{M_2}{M_1}\left(\frac{R_1}{D(t)}\right)^{l+1}W_{lm}\,e^{-i m f(t)}.
\end{equation}
We invoke von Zeipel's theorem \citep{vonzeipel24,pfahl08} to determine the corresponding surface emitted flux perturbation:
\begin{equation}\label{e:vZ}
 \frac{\Delta F_{lm}(t)}{F_1}=-(l+2)\frac{\xi_{r,lm}(t)}{R_1}.
\end{equation}

We can then explicitly evaluate the observed flux variation using the formalism from \se{s:obsflux}:
\begin{equation}\label{e:eqobsflux}
 \frac{\delta J}{J_1} = \frac{M_2}{M_1}\sum_{l=2}^\infty\left(\frac{R_1}{D(t)}\right)^{l+1}\sum_{m=-l}^l \Big\{\big[2-\beta(T_1)(l+2)\big] b_l- c_l\Big\}\,W_{lm}\,\Yo\, e^{-i m f(t)},
\end{equation}
where the bandpass correction coefficient $\bandpassT$ is defined in \eq{e:bandpass}, $W_{lm}$ is defined in \eq{e:W}, the disk-integral factors $b_l$ and $c_l$ are defined in equations \eqref{e:diskb} and \eqref{e:diskc},  several values of $b_l$ and $c_l$ using Eddington limb darkening are given in \ta{t:diskint}, and other variables are defined in \se{s:prelim}. Due to the strong dependence on $l$, it is typically acceptable to include only the first term of the sum in \eq{e:refl}. Computation of the binary separation $D(t)$ and true anomaly $f(t)$ is discussed in \ap{a:reflect}. The observed flux perturbation from the secondary is obtained from \eq{e:eqobsflux} by switching $1\leftrightarrow2$ and sending $\phi_o\rightarrow \phi_o+\pi$.

We note that although the analytic equilibrium tide solution for the radial displacement $\xi_r$ is a good approximation at the stellar surface regardless of stellar parameters, the presence of a significant surface convection zone in a solar-type star proscribes the use of \eq{e:vZ}; \citet{pfahl08} gives the appropriate replacement in their eq.~(37). Moreover, we  note that \eq{e:vZ} may also be invalid for slowly rotating stars in eccentric orbits; see \se{s:nonad}.

\section{Tidal Orbital Evolution}\label{a:sync}
\subsection{Eigenmode expansion of tidal torque and energy deposition rate}\label{a:torque}
Assuming alignment of rotational and orbital angular momenta, the tidal torque $\tau$ produced by star 2 on star 1 must have only a $z$ component, where $\uvec{z}$ points along the orbital angular momentum. We can evaluate it as follows \citep{kumar98}. First,
\newcommand{\dV}{\,dV}
\begin{equation}
 \begin{split}
  \tau &= \uvec{z}\cdot\int_* \left(\vec{r}\times\frac{d\vec{F}}{dV}\right) \dV = \int_* (\uvec{z}\times\vec{r})\cdot \frac{d\vec{F}}{dV}\,\dV\\
  &= \int_* \uvec{\phi}\cdot\left[-(\rho_0+\delta\rho)\vec{\nabla}U\right] r\sin\theta\dV.
 \end{split}
\end{equation}
The term involving the background density $\rho_0$ vanishes; expanding both the tidal potential $U$ and the Eulerian density perturbation $\delta\rho$ in spherical harmonics with expansion coefficients $U_{lm}(r,t)$ and $\delta\rho_{lm}(r,t)$, we have
\begin{equation}
 \tau(t) = i\qsphsum m\int_0^{R_1}\delta\rho_{lm}(r,t)U^*_{lm}(r,t)r^2dr.
\end{equation}
Further invoking the expansions from equations \eqref{e:summodes} and \eqref{e:Uexp}, as well as the definitions in \se{s:nmodes}, we arrive with
\begin{equation}
  \tau(t) = -2\,i\left(\frac{GM_1^2}{R_1}\right) \sum_{nlmkk'} m\frac{(\epsilon_l Q_{nl} W_{lm})^2}{ E_{nl}}\widetilde{X}^{k}_{lm}\widetilde{X}^{k'}_{lm}\lorentz e^{i(k'-k)\omegorb t}.
\end{equation}
Lastly, averaging over a complete orbital period and rearranging the sums, we derive our final expression for the secular tidal torque:
\newcommand{\avetau}{\langle\tau\rangle}
\begin{equation}\label{e:avetau}
  \avetau = 8\left(\frac{GM_1^2}{R_1}\right)\left(\frac{M_2}{M_1}\right)^2 \sum_{l=2}^\infty\left(\frac{R_1}{a}\right)^{2l+2} \sum_{m=-l}^l m W_{lm}^2 \sum_{k=0}^\infty \hans(e)^2\sum_{n} \left(\frac{Q_{nl}^2}{E_{nl}}\right)\left(\frac{\omegnl^2\dopp\gamma_{nl}}{(\omegnl^2-\dopp^2)^2+4\gamma_{nl}^2\dopp^2}\right).
\end{equation}
The torque depends on the rotation rate $\omegrot$ only through the Doppler-shifted frequency $\dopp = k\omegorb - m\omegrot$, since we have neglected rotational modification of the eigenmodes (\se{s:rot}). \fig{f:torque} shows plots of this torque evaluated for \koi{}.

Note that a particular term of this sum is positive if and only if $m\dopp=m(k\omegorb-m\omegrot)>0$, which reduces to $(k/m)\omegorb>\omegrot$. This is known as being prograde, since it is equivalent to the condition that a mode's angular structure, in the corotating frame, rotate in the same sense as the stellar spin; conversely, retrograde waves with $(k/m)\omegorb<\omegrot$ cause negative torques.

Using similar techniques to those given above, an equivalent expansion of the secular tidal energy deposition rate into the star (including mechanical rotational energy) can be derived:
\begin{equation}\label{e:aveEdot}
 \langle \dot{E}\rangle = 8\omegorb\left(\frac{GM_1^2}{R_1}\right)\left(\frac{M_2}{M_1}\right)^2 \sum_{l=2}^\infty\left(\frac{R_1}{a}\right)^{2l+2} \sum_{m=-l}^l W_{lm}^2 \sum_{k=0}^\infty k\hans(e)^2\sum_{n} \left(\frac{Q_{nl}^2}{E_{nl}}\right)\left(\frac{\omegnl^2\dopp\gamma_{nl}}{(\omegnl^2-\dopp^2)^2+4\gamma_{nl}^2\dopp^2}\right).
\end{equation}
The only difference between equations \eqref{e:avetau} and \eqref{e:aveEdot} is switching $m\leftrightarrow k\omegorb$.

\subsection{Nonresonant pseudosynchronization}\label{a:pseudo}
\newcommand{\avetaunr}{\langle\tau_\mr{nr}\rangle}
A pseudosynchronous frequency $\omegps$ is defined as a rotation rate that produces no average tidal torque on the star throughout a sufficiently long time interval, which here we take to be a complete orbital period (\se{s:sync}). I.e.,
\begin{equation}
\avetau(\omegps) = 0.
\end{equation}
Here we will show that our expansion from \ref{a:torque} reproduces the value of $\omegps$ derived in \citet{hut81}, which we denote $\omegpse$, in the equilibrium tide limit. We will in particular show that Hut's result is independent of assumptions about eigenmode damping rates.

Proceeding, we take the nonresonant (equilibrium tide) limit of \eq{e:avetau}. This is obtained by retaining only the first term in the Taylor series expansion in $\dopp/\omegnl$ of the last factor in parentheses from \eq{e:avetau}, and yields
\begin{equation}\label{e:avetaunr}
 \avetaunr = 8\left(\frac{GM_1^2}{R_1}\right)\left(\frac{M_2}{M_1}\right)^2 \sum_{l=2}^\infty\left(\frac{R_1}{a}\right)^{2l+2} \left[\sum_{m>0}^l m W_{lm}^2 \sum_{k=-\infty}^\infty \hans(e)^2\dopp\right]\left[\sum_{n} \left(\frac{Q_{nl}^2}{E_{nl}}\right)\left(\frac{\gamma_{nl}}{\omegnl^2}\right)\right];
\end{equation}
note that sums over $k$ and $m$ become decoupled from the sum over $n$. Setting $\avetaunr(\omegpse) = 0$ and retaining only $l=2$, we have
\begin{equation}\label{e:pseudodef}
 0=\sum_{k=-\infty}^\infty X_{22}^k(e)^2\left(k\omegorb-2\omegpse\right).
\end{equation} 

We need two identities to evaluate this further. First, starting with the definition of the Hansen coefficients,
\begin{equation}\label{e:hansdef}
 \left(\frac{a}{D}\right)^{l+1} e^{-i m f} = \sum_{k=-\infty}^\infty \hans e^{-ik\omegorb t},
\end{equation}
we can differentiate with respect to $t$, then multiply by the complex conjugate of \eqref{e:hansdef} and average over a complete period to derive
\begin{equation}
 \sum_{k=-\infty}^\infty k\left(\hans\right)^2 = \frac{m}{2\pi}\int_0^{2\pi}\left(\frac{1+e\cos f}{1-e^2}\right)^{2l+2}df.
\end{equation} 
Specializing to $l=2$, 
\begin{equation}\label{e:ident1}
 \sum_{k=-\infty}^\infty k\left(X_{2m}^k\right)^2 = m\left[\frac{5e^6+ 90e^4+120e^2 + 16}{16(1-e^2)^6}\right].
\end{equation}
The second identity needed,
\begin{equation}\label{e:ident2}
 \sum_{k=-\infty}^\infty \left(X_{2m}^k\right)^2 =\frac{3e^4+24e^2+8}{8(1-e^2)^{3/2}},
\end{equation}
can be derived similarly. 

Substituting equations \eqref{e:ident1} and \eqref{e:ident2} into \eqref{e:pseudodef}, we have that
\begin{equation}\label{e:hutpseudo}
 \omegpse = \omegorb\cdot\frac{1+(15/2)e^2+(45/8)e^4+(5/16)e^6}{\left[1+3e^2+(3/8)e^4\right](1-e^2)^{3/2}};
\end{equation} 
this is precisely eq.~(42) from \citet{hut81}.

\bibliography{ecctide}

\end{document}